\documentclass[12pt]{article}
  
\usepackage[margin=2.5cm]{geometry} 

\usepackage{amsmath}

\usepackage{amsthm}
\usepackage{amsbsy}
\usepackage{amssymb}
\usepackage[dvips]{graphicx}

\newcommand{\R}{\mathbb{R}}

\newcommand{\Z}{\mathbb{Z}}
\newcommand{\C}{\mathbb{C}}
\newcommand{\B}{\mathbb{B}}
\newcommand{\wpp}{\mathrm{wor-pro}}
\newcommand{\ap}{\mathrm{avg-pro}}
\newcommand{\pr}{\mathbf{p}}

\newcommand{\ve}{\varepsilon}

\newcommand{\lb}{\linebreak}
\newcommand{\br}[1]{|#1 \rangle}
\newcommand{\kt}[1]{\langle #1|}
\newcommand{\cl}[1]{\left\lceil #1 \right\rceil}
\newcommand{\fl}[1]{\left\lfloor #1 \right\rfloor}
\newcommand{\co}[1]{\overline{#1}}
\newcommand{\tha}{\arcsin\sqrt{a}}
\newcommand{\inn}[2]{\langle #1| #2 \rangle}
\newcommand{\x}{\overline{x}}
\newcommand{\y}{\underline{x}}

\title{Sharp Error Bounds on Quantum Boolean Summation \\
  in Various Settings\thanks{This research was supported in part by
    the National Science Foundation (NSF) and by the Defence Advanced
    Research Agency (DARPA) and Air Force Research Laboratory under
    agreement F30602-01-2-0523.}}

\author{\\
Marek Kwas and Henryk Wo\'zniakowski\\
\\
\\
Department of Computer Science \\
Columbia University\\
and\\
Institute of Applied Mathematics \\
University of Warsaw
}

\newtheorem{Brass&co}{Theorem} 
\newtheorem{PerfAnalys}[Brass&co]{Theorem}
\newtheorem{asymppp}[Brass&co]{Theorem}
\newtheorem{WA4}[Brass&co]{Theorem}
\newtheorem{WAn4}[Brass&co]{Theorem}

\newtheorem{MaxCor}{Corollary}
\newtheorem{MinCor}[MaxCor]{Corollary}
\newtheorem{IntegerCor}[MaxCor]{Corollary}
\newtheorem{ImprovedCor}[MaxCor]{Corollary}
\newtheorem{GlobalCor}[MaxCor]{Corollary}
\newtheorem{EpsilonCor}[MaxCor]{Corollary}

\begin{document}

\maketitle

\begin{abstract}
We study the quantum summation (\textbf{QS}) algorithm 
of Brassard, H\o{}yer, Mosca and Tapp, see
\cite{BMHT}, that approximates the arithmetic mean of a Boolean
function defined on $N$ elements. We improve error bounds 
presented in \cite{BMHT} in the worst-probabilistic setting,
and present new error bounds in the average-probabilistic setting. 

In particular, in the worst-probabilistic setting, we prove that the
error of the \textbf{QS} algorithm using $M-1$ quantum queries is
$\tfrac{3}{4}\pi M^{-1}$ with probability $\tfrac{8}{\pi^2}$, which
improves the error bound $\pi M^{-1} + \pi^2M^{-2}$ of \cite{BMHT}. We
also present error bounds with probabilities $p\in (\tfrac{1}{2},
\tfrac{8}{\pi^2}]$, and show that they are sharp for large $M$ and
$NM^{-1}$.

In the average-probabilistic setting, we prove that the \textbf{QS}
algorithm has error of order $\min \{M^{-1}, N^{-1/2}\}$  iff $M$ is
divisible by $4$. This  bound is optimal, as recently shown
in \cite{Papa}. For $M$ not divisible by $4$, the \textbf{QS} algorithm
is far from being optimal if $M \ll N^{1/2}$ since its error is
proportional to $M^{-1}$. 
\end{abstract}

\section{Introduction}
\label{sec:intro}
The quantum summation (\textbf{QS}) algorithm (also known as the
amplitude estimation algorithm) of Brassard, H\o{}yer, Mosca and Tapp,
see \cite{BMHT}, computes an approximation to the arithmetic mean of
all values of a Boolean function defined on a set of $N=2^n$
elements. Information regarding the Boolean function is supplied by
quantum queries.  The quantum queries play a role similar to the use
of function values in the worst case and randomized settings.  Suppose
that we use $M-1$ quantum queries. Obviously, the only case of
interest is when $M$ is much smaller than $N$.  It was proven in
\cite{BMHT} that the error of the \textbf{QS} algorithm is at most
\begin{equation}
\label{111}
\frac{\pi}M + \frac{\pi^2}{M^2}\quad \mbox{with probability}\
\frac8{\pi^2}=0.81\dots.  
\end{equation}

Nayak and Wu, see \cite{NayakWu}, showed that for any $p \in (
\tfrac{1}{2}, 1]$ the error of any quantum algorithm that uses no more
than $M-1$ quantum queries must be proportional to $M^{-1}$ with
probability $p$.  Therefore, the \textbf{QS} algorithm enjoys the
smallest possible error modulo a factor multiplying $M^{-1}$.

The minimal error estimate of order $M^{-1}$ in the quantum setting
should be compared to the minimal error estimates in the worst case
and randomized settings of algorithms using $M-1$ function
values. It is known, see \cite{N00}, that in the worst case setting,
the error bound is roughly $\tfrac{1}{2}(1-M/N)$. 
This means that as long as $M$
is much less than $N$ the error is almost $\tfrac{1}{2}$, and is 
therefore of order $M$
times larger than in the quantum setting. In the randomized setting,
the classical Monte Carlo is almost optimal, and the error bound is
roughly $1/(2\sqrt{M})$, see again \cite{N00}. Hence, it is of order
$\sqrt{M}$ larger than in the quantum setting.

The \textbf{QS} algorithm has many applications. In particular, it can be
used for approximation of the arithmetic mean of a real function, which is
the basic step for approximation of many continuous problems such as
multivariate integration, multivariate approximation and path integration,
see \cite{H1,H2,HN,NSW,TW}. 

Since the \textbf{QS} algorithm has so many applications, it seems
reasonable to check whether the estimate (\ref{111}) is sharp and how
the error decreases if we lower the probability $p=\tfrac{8}{\pi^2}$ to
$p>\tfrac{1}{2}$. It also seems reasonable to study the error of the
\textbf{QS} algorithm in various settings.  The estimate (\ref{111})
corresponds to the worst-probabilistic setting, which is most
frequently used in the quantum setting.  The essence of this setting
is that it holds for all Boolean functions.  It is also interesting to
study the average performance of the \textbf{QS} algorithm with respect
to some measure on Boolean functions.  This is the
average-probabilistic setting. In the worst-average and
average-average settings, we study the worst or average performance
with respect to Boolean functions and the average performance with
respect to all outcomes of a quantum algorithm. We add in passing that
the worst-average setting is usually used for the study of the
classical Monte Carlo algorithm. 

Sharp error bounds in the worst- and average-probabilistic settings
are addressed in this paper whereas the worst- and average-average
settings will be studied in a future paper. We study error bounds with
probabilities $p \in (\tfrac{1}{2}, \tfrac{8}{\pi^2}]$.  If we want to
obtain error bounds with higher probability, it is known that it is
enough to run the \textbf{QS} algorithm several times and take the
median as the final result, see e.g., \cite{H1}.

In the worst-probabilistic setting, we show that (\ref{111}) can be
slightly improved. Namely, the error of the \textbf{QS} algorithm is
at most
$$
\frac34\,\frac{\pi}M\quad\mbox{with probability}\ \frac8{\pi^2}.
$$ 
 
Furthermore, for large $M$ and $N/M$ we prove that the last estimate
is sharp. More generally, for $p\in(\tfrac{1}{2},\tfrac{8}{\pi^2}]$ 
we prove that the
error of the \textbf{QS} algorithm is at most
$$
\frac{(1-v^{-1}(p))\,\pi}{M}\qquad
\mbox{with probability} \ p,
$$
where $v^{-1}$ is the inverse of the function
$v(\Delta)=\sin^2(\pi\Delta)/ (\pi\Delta)^2$. 
We prove that the last estimate is sharp for large
$M$ and $N/M$. We have
$1-v^{-1}(p)\in(\tfrac{1}{2},\tfrac{3}{4}]$ and it is well approximated by
$\tfrac{1}{16}\pi^2p+\tfrac{1}{4}$. 
In particular, for the most commonly used values of $p$
we have
$$
(1-v^{-1}(\tfrac{1}{2}+))\pi=1.75\dots,\
(1-v^{-1}(\tfrac{3}{4}))\pi=2.23\dots,\
(1-v^{-1}(\tfrac{8}{\pi^2}))=\tfrac{3}{4}\pi=2.35\dots.
$$

In the average-probabilistic setting, we consider two measures on the
set of Boolean functions. The first measure is  uniform on Boolean
functions, while the second measure is  uniform on arithmetic means of
Boolean functions.  The results for these two measures are quite
different. The mean element of the arithmetic means is $\tfrac{1}{2}$ for both
measures. However, the first moment is of order $N^{-1/2}$ for the
first measure, and about $\tfrac{1}{4}$ for the second. The first moment is
exactly equal to the error of the constant algorithm that always outputs
$\tfrac{1}{2}$. This explains why we can obtain the
error of order $N^{-1/2}$ without any quantum queries
for the first measure. This provides
the motivation for us to check whether the error of the \textbf{QS}
algorithm enjoys a similar property. It turns out that this is indeed
the case iff $M$ is divisible by $4$. That is, for $M$ divisible by
$4$, the average-probabilistic error of the \textbf{QS} algorithm is
of order $\min\{M^{-1},N^{-1/2}\}$, and if $M$ is not divisible by
$4$, then the error is of order $M^{-1}$. For the second measure,
since the first moment is not small, the average-probabilistic
error of the \textbf{QS}
algorithm is of order $M^{-1}$ for all $M$.  For both measures, the upper
bounds presented in this paper match lower bounds that were recently
obtained by Papageorgiou, see~\cite{Papa}. Hence, the \textbf{QS}
algorithm enjoys minimal error bounds also in the
average-probabilistic setting if we choose $M$ divisible by $4$
for the first measure and with no restriction on $M$ for the second measure.

The quantum setting, and in particular the \textbf{QS} algorithm, is
relatively new and probably not well known, especially for people
interested in continuous complexity. Hence we present all details of
this algorithm, emphasizing its quantum parts.  Since we wanted also
to find sharp error bounds, we needed a very detailed analysis of the
outcome probabilities of the \textbf{QS} algorithm.

We outline the contents of this paper. In Section
\ref{sec:QuantSummAlgo} we define the \textbf{QS} algorithm. Section
\ref{sec:PerfAnalys} deals with the performance analysis of the
\textbf{QS} algorithm in the worst-probabilistic setting, see Section
\ref{sec:WCP}, and in the average-probabilistic setting, see
Section~\ref{sec:ACP}.

\section{Quantum Summation Algorithm}
\label{sec:QuantSummAlgo}

We consider the most basic form
of the summation problem, i.e., the summation of Boolean
functions. Let $\B_N$ denote the set of Boolean functions $f:\{0,
 \ldots, N-1\} \rightarrow \{0,1\}$. Let
\begin{equation*}
  a_f=\frac{1}{N}\sum_{i=0}^{N-1} f(i)
\end{equation*}
denote the arithmetic mean of all values of $f$. Clearly, $a_f\in[0,1]$.

\vskip 1pc
\noindent \textbf{Problem:} For $f\in \B_N$, compute an  $\ve$-approximation 
$\bar a_f$ of the sum $a_f$ such that 
\begin{equation}
\label{eq:sum_app}
  |\bar a_f - a_f| \le \ve.
\end{equation}
\vskip 1pc

We are interested in the minimal number of evaluations of the function
$f$ that are needed to compute $\bar a_f$
satisfying~\eqref{eq:sum_app}. It is known that in the worst case
setting, we need roughly $N(1-\ve)$ evaluations of the function $f$. 
In the randomized setting, we assume that $\bar a_f$ is a
random variable and require that \eqref{eq:sum_app} 
holds for the expected value of
$| \bar a_f - a_f|$ for any function~$f$. It is known, see
e.g., \cite{N00}, that in the randomized setting we need roughly
$\min\{N, \ve^{-1/2}\}$ function evaluations. In the quantum setting,
we want to compute a random variable $\bar a_f$ such that  
\eqref{eq:sum_app} holds with a high probability (greater than 
$\tfrac{1}{2}$) either for \emph{all} Boolean functions or 
on the average with respect to a
probability measure defined on the set $\B_N$. These two error
criteria in the quantum setting will be precisely defined in
Section~\ref{sec:PerfAnalys}.

In this section we describe the quantum summation algorithm, which is
also called the quantum amplitude estimation algorithm. This algorithm
was discovered by Brassard, H\o{}yer, Mosca and Tapp \cite{BMHT}, and uses
Grover's iterate operator as its basic component, see \cite{G98}.  
We use standard notation of quantum computation, see e.g., \cite{NC}. 

For simplicity we assume that $N=2^n$.  Let $\mathcal{H}_n$ denote the
tensor product $\C^2 \otimes \cdots \otimes \C^2$ of $n$ copies of
$\C^2$, with $\C^2$ the 2-dimensional complex vector space. Unit
vectors from $\C^2$ are called \emph{one qubit} quantum states (or
\emph{qubits}). Let 
$\br{0}$ and $\br{1}$ be an orthonormal basis of $\C^2$. Then any
qubit~$\br{\psi}$ can be represented as
$$
\br{\psi}=\psi_0 \br{0}+
\psi_1 \br{1}\qquad \text{with}\quad \psi_k\in \C \quad\text{and}\quad|\psi_0|^2+|\psi_1|^2=1.
$$
For $j=0,1,\ldots, N-1$, we have $j=\sum_{k=0}^{n-1}2^{n-1-k} j_k$, with $j_k
\in \{0,1\}$. Let 
$$
\br{j}=\bigotimes_{k=0}^{n-1}\br{j_k}.
$$  
The set
$\{\br{j}:j=0,\ldots,N-1\}$ forms an orthonormal basis of
$\mathcal{H}_n$ and any unit vector $\br{\psi}\in \mathcal{H}_n$ can
be represented as 
$$
\br{\psi}=\sum_{j=0}^{N-1}\psi_j\br{j}\qquad \text{with}\quad \psi_j
\in \C \quad \text{and}\quad\sum_{j=0}^{N-1} |\psi_j|^2 =1.
$$
Unit vectors from $\mathcal{H}_n$ are called 
$n$ qubit quantum states (or quantum states or just states,
whenever $n$ is clear from the context).

The only transforms that can be performed on
quantum states are defined by certain unitary operators on
$\mathcal{H}_n$.   
We now define the six unitary operators that are basic components of the
summation algorithm. Since unitary operators are linear, it is enough
to define them on the basis states~$\br{j}$.
\begin{enumerate}

\item Let $S_0:\mathcal{H}_n\rightarrow\mathcal{H}_n$ 
denote the \emph{inversion about zero transform}
  \begin{equation*}
    S_0 \br{j}=(-1)^{\delta_{j,0}}\br{j},
  \end{equation*}
where $\delta_{j,0}$ is the Kronecker delta. Hence, $S_0 \br{0}=-\br{0}$
and $S_0 \br{j}=\br{j}$ for all $j\ne 0$. This corresponds to the
diagonal matrix with one element equal to $-1$, and the rest equal to
$1$. The operator $S_0$ can be also written as the Householder operator
\begin{equation*}
  S_0=I-2\br{0}\kt{0}.
\end{equation*}
Here, for a state $\br{\psi}$, we let 
$\br{\psi}\kt{\psi}$ denote the projection onto the space
$\mathrm{span}\{\br{\psi}\}$ given by 
\begin{equation*}
  (\br{\psi}\kt{\psi})\,\br{x}= \kt{\psi}x\rangle\,\br{\psi},
\end{equation*}
where $\kt{\psi}x\rangle$ is the inner product\footnote{We follow the
  quantum mechanics notation in which the first argument is
  conjugated in the inner product, whereas in the standard
  mathematical notation the second argument is usually conjugated.}  
in $\mathcal{H}_n$, $\kt{\psi} x\rangle=\sum_{k=0}^{N-1}  \co{\psi_k} 
x_k$. The matrix form of the projector $\br{\psi}\kt{\psi}$ in the
basis $\{\br{j}\}$ is $( \co{\psi_k} \psi_j)_{j,k=0}^{N-1}$. One  can
also view the matrix form of 
the projector $\br{\psi}\kt{\psi}$ as the matrix product of 
the $N\times 1$ column
vector $\br{\psi}$ and the $N\times 1$ row vector 
$\kt{\psi}$, which 
is the Hermitian conjugate of $\br{\psi}$, $\kt{\psi}= \br{\psi}^\dag$.
For any $\br{x}\in \mathcal{H}_n$ we have 
\begin{equation*}
\kt{k}\big(I-2\br{0}\kt{0}\big)\br{x}=\inn{k}{x}-2\inn{0}{x}\inn{k}{0}=
\begin{cases}
x_k -2x_k =-x_k &\text{for}\; k=0,\\
x_k-0=x_k  &\text{for} \;k \ne 0.
\end{cases}
\end{equation*}
Hence $I-2\br{0}\kt{0}=S_0$, as claimed.

\item  Let $W_N:\mathcal{H}_n\rightarrow\mathcal{H}_n$ denote the 
\emph{Walsh-Hadamard transform} 
  \begin{equation*}
    W_N \br{j}= 
\frac{1}{\sqrt{N}}
    \bigotimes_{k=0}^{n-1}\left(\br{0}+(-1)^{j_k}\br{1}\right). 
  \end{equation*}
That is, the Walsh-Hadamard transform corresponds to the matrix
with entries 
\begin{equation*}
  \kt{i} W_N \br{j} = \frac{1}{\sqrt{N}} \prod_{k=0}^{n-1}
  \kt{i_k}\left(\br{0}+(-1)^{j_k}\br{1}\right)= \frac{1}{\sqrt{N}}
  \prod_{k=0}^{n-1} (-1)^{i_k j_k} =\frac{1}{\sqrt{N}}
  (-1)^{\sum_{k=0}^{n-1}i_kj_k}.
\end{equation*}
The matrix $(\kt{i} W_{N} \br{j})_{i,j=0}^{N-1}$ is symmetric.
Furthermore,
\begin{eqnarray*}
W_N^2 \br{j}\,&=&\, \frac{1}{\sqrt{N}}\,W_n 
    \bigotimes_{k=0}^{n-1}\left(\br{0}+(-1)^{j_k}\br{1}\right)\\
&=&\,\frac{1}{\sqrt{N}}\bigotimes_{k=0}^{n-1}\left(
\frac1{\sqrt{2}}\left(\br{0}+\br{1}\right)+\frac{(-1)^{j_k}}{\sqrt{2}}
\left(\br{0}-\br{1}\right)\right)\\
&=&\,\frac1{\sqrt{N}}\bigotimes_{k=0}^{n-1}\sqrt{2}\br{j_k}\,=\,\br{j}.
\end{eqnarray*}
Thus, $W_N^2=I$ and $W_N^{-1}=W_N$ is orthogonal. This means that
the operator $W_N$ is symmetric and unitary.
 
\item For $K=1,2,\ldots, 2^n$,  
let $F_{K,n}:\mathcal{H}_n\rightarrow\mathcal{H}_n$ denote the
\emph{ quantum Fourier transform} 
\begin{equation*}
  F_{K,n} \br{j}=
\begin{cases}
\frac{1}{\sqrt{K}}\sum_{k=0}^{K-1}e^{2 \pi i j
    k/K}\br{k}, &\text{for}\quad j=0,1,\ldots,K-1, \quad
(i=\sqrt{-1})\\
\br{j}  &\text{for} \quad j=K,\ldots,2^n-1.
\end{cases}
\end{equation*}
Hence, $F_{K,n}$ corresponds to the unitary block-diagonal matrix 
\begin{equation*} 
\left[
\begin{array}{cc}
F_K & 0\\
0 & I
\end{array}
\right],
\end{equation*} 
where
$F_K= \left(K^{-1/2}\,e^{2 \pi i j k /k}\right)_{j,k=0}^{K-1}$ is the
matrix of the inverse  quantum Fourier transform. 
For $K=2^n=N$ we have 
\begin{equation*}
  F_{N,n} \br{\psi}=\sum_{j=0}^{N-1} \psi_j F_{N,n} \br{j}= \frac{1}{\sqrt{N}}
  \sum_{k=0}^{N-1} \left(\sum_{j=0}^{N-1} \psi_j 
  e^{2\pi i j k/N}\right) \br{k}.
\end{equation*}
The coefficients of $F_{N,n} \br{\psi}$ in the basis $\{\br{j}\}$ are
the inverse quantum Fourier transforms of the coefficients of the
state~$\br{\psi}$.  Note that $W_N$ and $F_{N,n}$ coincide for the
state $\br{0}$, i.e.,
$$
W_N\,\br{0}\,=\,F_{N,n}\,\br{0}\,=\,\frac1{\sqrt{N}}\sum_{j=0}^{N-1}\br{j}.
$$

\item 
Let $S_f:\mathcal{H}_n\rightarrow\mathcal{H}_n$ 
denote the \emph{quantum query} operator
\begin{equation*}
 S_f \br{j} = (-1)^{f(j)} \br{j},
\end{equation*}
This again corresponds to the diagonal matrix with elements $\pm 1$
depending on the values of the Boolean function $f$. 
This operator is the only one that provides information about the
Boolean function $f$. This is analogous 
to the concept of  an \emph{oracle} or a
\emph{black-box} which is used in classical computation and which
supplies information about the function  $f$ through its  values.

The standard definition of the quantum query $\bar S_f$ is 
\begin{equation*}
  \bar S_f: \mathcal{H}_n\otimes \C^2
  \rightarrow\mathcal{H}_n\otimes\C^2,\qquad  \bar S_f
  \br{j}\br{i}=\br{j}\br{i \oplus f(j)},
\end{equation*}
where $\oplus$ means addition modulo 2. We can simulate
$S_f$ by $\bar S_f$ if we use an auxiliary qubit
$(1/\sqrt{2})(\br{1}-\br{0})$, namely,
\begin{multline*}
  \bar S_f\,
  \bigg(\br{j}\frac{\br{1}-\br{0}}{\sqrt{2}}\bigg)=\br{j}\frac{\br{1\oplus
  f(j)}-\br{f(j)}}{\sqrt{2}} \\ =(-1)^{f(j)}\br{j}\frac{\br{1}-\br{0}}{\sqrt{2}}
  = \big(S_f\br{j}\big)\frac{\br{1}-\br{0}}{\sqrt{2}}.
\end{multline*}

\item Let $Q_f: \mathcal{H}_n \rightarrow\mathcal{H}_n$ 
denote the \emph{Grover operator}
  \begin{equation*}
    Q_f=-W_N\,S_0\,W_N^{-1}\, S_f.
  \end{equation*}
  This is the basic component of Grover's search algorithm, see \cite{G98}.
  As we shall see, $Q_f$ also plays the major role for the summation
  algorithm. The eigenvectors and eigenvalues of $Q_f$
  will be useful in further considerations. Let
$$
\br{\psi}=W_N \br{0} =\frac{1}{\sqrt{N}}\sum_{k=0}^{N-1} \br{k}
$$
 and $\br{\psi_0}$, $\br{\psi_1}$
denote the orthogonal projections of $\br{\psi}$ onto the subspaces \lb
$\mathrm{span}\{\br{j}:f(j)=0\}$ and $\mathrm{span}\{\br{j}:f(j)=1\}$,
respectively. That is, 
$$
\br{\psi_j}=\frac{1}{\sqrt{N}}
\sum_{k:\,f(k)=j} \br{k}\qquad j=0,1.
$$ 
Then 
$\br{\psi}=\br{\psi_0}+\br{\psi_1}$ and
$\inn{\psi_0}{\psi_1}=0$. 
Furthermore, 
$\inn{\psi_j}{\psi_j}= N^{-1}\sum_{k:\, f(k)=j} 1$, for $j=0,1$, 
so that $\inn{\psi_1}{\psi_1}=a$ and $\inn{\psi_0}{\psi_0}=1-a$,
where $a=a_f$ is the sum we want to approximate. 

{}From~\cite{BMHT}, we know  that
\begin{equation}
\label{Q}
\begin{split}
  Q_f\br{\psi_0}&=(1-2a)\br{\psi_0}+2(1-a)\br{\psi_1},\\
  Q_f\br{\psi_1}&=-2a\br{\psi_0}+(1-2a)\br{\psi_1}.
\end{split}
\end{equation}

For the sake of completeness, we provide a short proof of \eqref{Q}.
By the definition of the operator $S_f$ we have
$$
S_f\br{\psi_j}=(-1)^{j} \br{\psi_j},\quad j=0,1,
$$ 
and
$$
W_N\,S_0\,W_N^{-1}\, =\, W_N(I-2\br{0}\kt{0})\,W_N^{-1}
\,=\, I - 2 (W_N\br{0}\kt{0}W_N ).
$$
Since $\kt{0}W_N=(W_N\br{0})^\dag= (\br{\psi})^\dag
=\kt{\psi}$, we obtain for $j=0,1$, 
\begin{align*}
W_N\, S_0\, W_N^{-1} \br{\psi_j}= \br{\psi_j} -2
(\br{\psi}\kt{\psi})\,\br{\psi_j}
= \br{\psi_j}-2\inn{\psi}{\psi_j} \br{\psi}
=\br{\psi_j}-2 \inn{\psi_j}{\psi_j} \br{\psi}.
\end{align*}
{}From this we calculate for $j=0,1$, 
\begin{multline*}
Q_f\br{\psi_j}=(-1)^{1+j}W_NS_0W_N^{-1}\br{\psi_j} \\ =
(-1)^{\delta{j,0}} \left(\br{\psi_j} - 2
(\delta_{j,1}a+\delta_{j,0}(1-a))(\br{\psi_0}+\br{\psi_1})\right), 
\end{multline*}
which is equivalent to \eqref{Q}.   

Thus, the space $\mathrm{span}\{\br{\psi_0}, \br{\psi_1}\}$ is an 
invariant space of $Q_f$ and its eigenvectors and
corresponding eigenvalues can be computed by solving the eigenproblem
for the $2\times 2$ matrix
\begin{equation*}
  \left[
\begin{array}{cc}
1-2a & -2a\\
2(1-a) & 1-2a
\end{array}\right]\,.
\end{equation*}
For $a\in(0,1)$, the
eigenvectors and the corresponding orthonormalized eigenvalues 
of $Q_f$ are
\begin{equation*}
    \br{\psi_\pm}=\frac{1}{\sqrt{2}}\left(\pm\frac{i}{\sqrt{1-a}}
      \br{\psi_0}+ \frac{1}{\sqrt{a}}\br{\psi_1}\right)\quad
    \mathrm{and} \quad
    \lambda_{\pm}=1-2a\pm 2i\sqrt{a(1-a)}=e^{\pm 2i\theta_a},    \end{equation*} where $\theta_a=\arcsin\sqrt{a}$. 
Moreover, it is easy to check that
\begin{align}
\label{eq:represent}
  \br{\psi}=\frac{-i}{\sqrt{2}}\left(e^{i\theta_a}\br{\psi_+}-e^{-i\theta_a}\br{\psi_-}\right). 
\end{align}
For $a\in\{0,1\}$, we have
$\mathrm{span}\{\br{\psi_0}, \br{\psi_1}\}=
\mathrm{span}\{\br{\psi}\}$ and $\br{\psi}$ is the eigenvector of $Q_f$
with eigenvalues $\pm 1$, respectively. For $a\in\{0,1\}$, we define
$$
\br{\psi_+}\,=\,i^{1-a}\,\sqrt{2}\,\br{\psi}\quad\mbox{and}\quad
\br{\psi_{-}}=0.
$$
Then it is easy to check that (\ref{eq:represent}) is valid, and
$\lambda_{\pm}=e^{\pm 2i\theta_a}=(-1)^a$ is an eigenvalue of $Q_f$
for all $a\in[0,1]$.

\item The next unitary transform, called 
the \emph{Grover iterate operator}, is defined on the tensor product of 
$\mathcal{H}_m \otimes \mathcal{H}_n$ and uses $m+n$ qubits.
The first space $\mathcal{H}_m$ and $m$ qubits will be related to
the accuracy of the quantum summation algorithm.  
The \emph{Grover iterate} operator 
$\Lambda_m(Q_f): \mathcal{H}_m \otimes \mathcal{H}_n
\rightarrow \mathcal{H}_m \otimes \mathcal{H}_n$ is defined by
\begin{equation*}
\Lambda_m(Q_f)\,\br{j}\br{y} = \br{j}\, Q_f^{\,j} \br{y}\quad
\text{for}\quad \br{j}\br{y}\in \mathcal{H}_m\otimes \mathcal{H}_n.   
\end{equation*} 
Hence, the power of $Q_f$ applied to the second component depends on the
first one. Note that $j$ may vary from 0 to $2^m-1$. Therefore
$\Lambda_m(Q_f)$ may use the powers of $Q_f$ up to the $(2^m-1)$st.
\end{enumerate}
\vspace{12pt}

We need one more concept of quantum computation, that 
of \emph{measurement}. Suppose $s$ is a positive integer and
consider the space $\mathcal{H}_s$. Given the state  
$$
\br{\psi}=\sum_{k=0}^{2^s-1}\psi_k \br{k}\,\in\,\mathcal{H}_s,
$$
we cannot, in general, recover all the coefficients $\psi_k$.  
We can only measure the state $\br{\psi}$  with respect to a finite
collection of linear operators 
$\{M_j\}_{j=0}^p$, where the $M_j:\mathcal{H}_s\rightarrow\mathcal{H}_s$ 
satisfy the completeness relation 
$$
\sum_{j=0}^p M_j^\dag M_j =I.
$$ 
After performing the measurement, we obtain the outcome $j$
and the state $\br{\psi}$ collapses into the state 
$$
\frac{1}{\sqrt{\langle \psi|M_j^\dag M_j \br{\psi}}}\, M_j\br{\psi},
$$
both these events occur with probability $\langle
\psi|M_j^\dag M_j \br{\psi}$. Note that for $M_j\br{\psi}=0$ 
the outcome $j$ cannot happen with positive probability. 
Hence, with probability $1$ the outcome $j$ 
corresponds to $M_j\br{\psi}\not=0$ for $j=0,1,\ldots, p$. 

The most important example of such a collection of operators is
$\{\br{j}\kt{j}\}_{j=0}^{2^s-1}$. Then, the measurement of the state
$\br{\psi}$ with respect to this collection of operators gives us the
outcome~$j$ and the collapse into the state
$$\frac{\inn{j}{\psi}}{|\inn{j}{\psi}|}\br{j}
$$ 
with probability $|\psi_j|^2$, $j=0,1,\ldots,2^s-1$.


Another example is a variation of the 
previous example and will be used in the quantum summation algorithm. 
We now let $s=m+n$, as for the Grover iterate operator, and define
$M_j: \mathcal{H}_m\otimes \mathcal{H}_n \rightarrow \mathcal{H}_m \otimes
\mathcal{H}_n$ by 
$$M_j\,=\,\br{j}\kt{j}\otimes I
$$ for $j=0,1,\ldots,2^m-1$, 
with $I$ denoting the identity operator on $\mathcal{H}_n$. That is,
$$
\left(\br{j}\kt{j}\otimes I \right)\,\br{x}\br{y}\,=\,
\inn{j}{x}\,\br{j} \br{y}
$$ 
for $\br{x}\in\mathcal{H}_m$ and $\br{y}\in\mathcal{H}_n$.

Since $\sum_{j=0}^{2^m-1}\left(\br{j}\kt{j}\otimes I\right)\br{x}\br{y}
=\br{x}\br{y}$
for all basis states $\br{x}$ of $\mathcal{H}_m$ and $\br{y}$
of $\mathcal{H}_n$, the completeness relation 
is satisfied. Consider now the probability of the outcome $j$ for
a special state $\br{\psi}$ of the form $\br{\psi}=\br{\psi_1}\br{\psi_2}$
with $\br{\psi_1}\in\mathcal{H}_m$, $\br{\psi_2}\in\mathcal{H}_n$ and
$\inn{\psi_k}{\psi_k}=1$ for $k=1,2$. Since $\br{j}\kt{j}\otimes I$
is self-adjoint, the outcome $j$ and the collapse of the state $\br{\psi}$
to the state 
$$
\frac{\langle j \br{\psi_1}}{|\langle j \br{\psi_1}|}\,\br{j}\br{\psi_2}
$$
occur with probability $|\inn{j}{\psi_1}|^2$. 
Hence, this collection of operators measures the
components of the so-called first register $\br{\psi_1}$ of the quantum state
$\br{\psi}$.

\vspace{12pt}

Following \cite{BMHT}, we are ready to describe the quantum summation
(\textbf{QS}) algorithm for solving our problem. 
The \textbf{QS} algorithm depends on a Boolean function $f$ and on an
integer parameter $M$ that controls the number of quantum queries of $f$
used by the algorithm. We perform computations
in the space $\mathcal{H}_m \otimes \mathcal{H}_n$, with
$m=\cl{\log_2 M}$, so we use $n+m$ qubits. 
As we will see later, the accuracy of the
algorithm is related to the dimension of the space $\mathcal{H}_m$.  
\vspace{12pt}

\noindent \textbf{Algorithm {QS}}($f$, $M$)
\begin{description}
\item[Input state:] $\br{0}\br{0} \in \mathcal{H}_m\otimes \mathcal{H}_n$
with $m=\cl{\log_2 M}$ and $n=\log_2N$.
\item[Computation:]   \qquad
  \begin{enumerate}
  \item $\br{\eta_1}=F_{M,m}\otimes W_N\,\br{0}\br{0}$,
  \item $\br{\eta_2}=\Lambda_m(Q_f)\,\br{\eta_1}$,
  \item $\br{\eta_3}=(F_{M,m}^{-1} \otimes I)\, \br{\eta_2}$.
  \end{enumerate}
\item[Measurement:] \quad \lb Perform the measurement of the state
  $\br{\eta_3}$ with respect to the collection \lb 
  $\{\,(\br{j}\kt{j})\otimes I\,\}_{j=0}^{2^m-1}$. Denote the outcome
  by $j$.
\item [Output:] $\bar a_f(j) =\sin^2\big(\pi j /M\big)$.
\end{description}
\vspace{12pt}

We briefly comment on the \textbf{QS} algorithm. The input state is
always the same and does not depend on $f$. Step 1 computes
$\br{\eta_1}=(NM)^{-1/2} \sum_{j=0}^{M-1} \sum_{k=0}^{N-1} \br{j}
\br{k}$, which is the equally weighted superposition of the basis
states. Step 2 computes $\br{\eta_2}$ by using the Grover iterate operator.
During this step we use the successive powers of the Grover operator $Q_f$,
and this is the only step where information about the Boolean
function $f$ is used. We shall see that the \textbf{QS} algorithm uses $M -1$
quantum queries. Step~3 computes $\br{\eta_3}$
by performing the inverse quantum Fourier  transform on the first $m$
qubits, and  prepares the system for measurement. 
After Step 3, we perform the measurement, obtain the outcome $j$ 
and compute the output $\bar a_f(j)$ 
on a classical computer. We stress that 
the distribution of the outcomes $j$ depends on the Boolean function $f$,
and this is the only dependence of the output $\bar a_f(j)$ on~$f$.
 
\section{Performance Analysis}
\label{sec:PerfAnalys}
In this section we analyze the error of the \textbf{QS} algorithm. 
As we have seen in Section~\ref{sec:QuantSummAlgo}, the output 
$\bar a_f(j)$ of the \textbf{QS} algorithm 
is a random value chosen according to a
certain distribution dependent on the input function $f$. 
In this way, the \textbf{QS} algorithm is a randomized
algorithm. Various ways of measuring the performance of
randomized algorithms are commonly used in the analysis of
algorithms and computational complexity. They correspond to various
error criteria. In this paper we consider two error criteria:
 worst-probabilistic and average-probabilistic. In a future
paper we consider other two error criteria:
worst-average and average-average, which correspond to the
worst or average performance with respect to Boolean functions
and the average performance with respect to all outcomes.

\subsubsection*{Worst-Probabilistic Error}
We start with the error criterion that is used in most papers dealing
with quantum computations. We are interested in the worst case
error of the \textbf{QS} algorithm that holds with a
given probability $p$. Here $p \in [0,1]$ and $1-p$ measures the
probability of \textbf{QS} algorithm's failure and usually $p$
is set to be $\tfrac{3}{4}$. In our analysis, however, we will allow 
an arbitrary $p\in (\tfrac{1}{2},\tfrac{8}{\pi^2}]$. 
The choice of the upper bound
$\tfrac{8}{\pi^2}=0.81\dots$ will be clear from the analysis of
the \textbf{QS} algorithm. 
The \textbf{QS} algorithm outputs $\bar a_f(j)$ with probability
$p_f(j)$ for $j=0,1,\ldots, M-1$, see Theorem \ref{thm:PerfAnalys}
where the $p_j(f)$'s are given. 
Its worst-probabilistic error is formally defined as
the smallest error bound that holds for all Boolean function
with probability at least $p$, i.e., 
\begin{equation*}
  e^{\wpp}(M,p)=\inf \bigg\{\alpha: \sum_{j:\;|a_f-\bar a_f(j)|\le \alpha}
  p_f(j) \ge p \qquad \forall f \in \B_N\bigg\}.
\end{equation*}
It is easy to see that $e^{\wpp}(M,p)$ can be rewritten as
follows. Let $A \subset \{0,1,\ldots, M-1\}$. For $f \in \B_N$ define the
measure of $A$ as
\begin{equation*}
  \mu(A,f)=\sum_{j\in A} p_f(j).
\end{equation*}
Then

\begin{equation}
  \label{eq:wperr1}
  e^{\wpp}(M,p)=
\max_{f \in \B_N} \;\min_{A:\; \mu(A,f) \ge p}\; 
\max_{j\in A} |a_f -\bar  a_f(j)|.
\end{equation}

\subsubsection*{Average-Probabilistic Error}

The worst-probabilistic error $e^{\wpp}(M,p)$ of the \textbf{QS}
algorithm is defined by the worst performance with respect to Boolean
functions. It is also natural to consider the average performance of
the \textbf{QS} algorithm with respect to Boolean functions. Let $\pr$
be a probability measure on the set $\B_N$. That is, any Boolean
function $f\in \B_N$ occurs with probability $\pr(f)$. Obviously,
$\pr(f) \ge 0$ and $\sum_{f \in \B_N} \pr(f) =1$. The
average-probabilistic error is defined by replacing the first max in
\eqref{eq:wperr1} by the expectation, i.e., 
\begin{equation*}
e^{\ap}(M,p)=\sum_{f \in \B_N} \pr(f) \min_{A:\;
\mu(A,f) \ge p} \max_{j\in A} |a_f -\bar a_f(j)|, 
\end{equation*}
Hence, we are interested in the average error that
holds with a certain fixed probability.

\vspace{24pt}



\subsection{Worst-Probabilistic Error}
\label{sec:WCP}

We begin by citing a theorem from~\cite{BMHT} for which we will propose a
number of improvements.
\begin{Brass&co} 
\label{thm:Brass&co}
\cite{BMHT} For any Boolean function $f\in \B_N$, the \textbf{QS} algorithm 
uses exactly $M-1$ quantum queries and outputs $\bar a$
that approximates $a=a_f$  such that 
\begin{equation*}
  |\bar a - a| \le \frac{2 \pi}{M} \sqrt{a(1-a)} + \frac{\pi^2}{M^2}
\le \frac{\pi}M+\frac{\pi^2}{M^2}
\end{equation*}
with probability at least $\tfrac{8}{\pi^2}=0.81\ldots$ .
Hence,
\begin{equation*}
  \sum_{j:\; |\bar a_f (j) - a_f| \le (2\pi/M)
\sqrt{a_f(1-a_f)} + \pi^2/M^2} p_f(j)
  \ge \frac{8}{\pi^2}\qquad \forall f\in \B_N, 
\end{equation*}
and, therefore,  
\begin{equation*}
  e^{\wpp}(M,\tfrac{8}{\pi^2}) \le \frac{\pi}{M} + \frac{\pi^2}{M^2}.
\end{equation*}
\end{Brass&co}

Using proof ideas of Theorem \ref{thm:Brass&co} from \cite{BMHT} we
present the following theorem and the subsequent corollaries. 
\begin{PerfAnalys}
\label{thm:PerfAnalys}
For any Boolean function $f\in\B_N$, denote 
$$\sigma_a\,=\,\sigma_{a_f}\,=\,
\frac{M}{\pi} \arcsin\sqrt{a}\in \left[0,\tfrac{1}{2}M\right].
$$ 
\begin{enumerate}
\item The \textbf{QS} algorithm uses exactly $M-1$ quantum queries,
and $\log_2N\,+\cl{\log_2M}$ qubits.

\item
For $j=0,1,\ldots,M-1$, the outcome $j$ of 
the \textbf{QS} algorithm occurs with
probability 
\begin{equation}
  \label{eq:pj}
p_f(j)\,=\, \frac{\sin^2(\pi(j-\sigma_{a_f}))}{2M^2\sin^2(
\frac{\pi}{M}(j-\sigma_{a_f}))}
\left(1+\frac{\sin^2\left(\pi(j-\sigma_{a_f})/M\right)}
{\sin^2\left(\pi(j+\sigma_{a_f})/M\right)}\right).
\end{equation}
(If $\sin(\pi(j\pm \sigma_{a_f})/M)=0$ we need to apply the limiting
value of the formula above.)
For $j=M, M+1,\ldots,2^{\cl{\log_2 M}}-1$, the outcome $j$ occurs with
probability $0$.  

\item If $\sigma_{a_f}$ is an integer then the \textbf{QS} algorithm outputs
the exact value of $a_f$ with probability~$1$. This holds iff 
$a_f=\sin^2(k\pi/M)$ for some integer $k\in[0,\tfrac{1}{2}M]$. 
In particular, this
holds for $a_f=0$, for $a_f=1$ and even $M$, and for $a_f=\tfrac{1}{2}$ and
$M$ divisible by $4$. 

\item Let $\x= \pi(\cl{\sigma_a}-\sigma_a)/M$ and
$\y=\pi(\sigma_a-\fl{\sigma_a})/M$. 
If $\sigma_{a_f}$ is not an integer then the \textbf{QS} algorithm
outputs the same value $ \bar a=\bar a_f(\cl{\sigma_a})=
\bar a_f(M-\cl{\sigma_a})$ for the outcomes
$\cl{\sigma_a}$ and $M-\cl{\sigma_a}$ such that
\begin{eqnarray}
  \label{eq:perfAnalysErr1}
  |\bar a - a|\,&=&\,\left|\sin (\x)
\left(2\sqrt{a(1-a)}\cos(\x)
+(1-2a)\sin(\x)\right)\right| 
\nonumber \\
&\le&\,\frac{\pi}{M}\big(\cl{\sigma_a} - \sigma_a\big)
\end{eqnarray}
with probability 
\begin{eqnarray}
\label{eq:perfAnalysPr1}
&& \frac{\sin^2 (\pi(\cl{\sigma_a} - \sigma_a))}{M^2 \sin^2
    (\frac{\pi}{M}(\cl{\sigma_a} - \sigma_a))}
\,\left(1+(1-\delta_{\cl{\sigma_a},M/2})
\frac{\sin^2(\pi(\cl{\sigma_a}-\sigma_a)/M)}{\sin^2(\pi(\cl{\sigma_a}+
\sigma_a)/M)}\right) \nonumber  \\
&\ge&
   \frac{\sin^2
    \big(\pi(\cl{\sigma_a} - \sigma_a)\big)}{\pi^2(\cl{\sigma_a} -
    \sigma_a)^2}= 1- \frac{\pi^2}{3}(\cl{\sigma_a} - \sigma_a)^2 +O
     \big((\cl{\sigma_a} - \sigma_a)^4\big), 
\end{eqnarray}
and outputs the same value $\bar a=\bar a_f(\fl{\sigma_a})=
\bar a_f((1-\delta_{\fl{\sigma_a},0})M-\fl{\sigma_a})$ for the outcomes
$\fl{\sigma_a}$ and $(1-\delta_{\fl{\sigma_a},0})M-\fl{\sigma_a}$such that 
\begin{eqnarray}
  \label{eq:perfAnalysErr2}
 |\bar a - a|\,&=&\,\left|\sin(\y)
\left(2\sqrt{a(1-a)}\cos(\y)
+(1-2a)\sin(\y)\right)\right| 
\nonumber \\
&\le&\, \frac{\pi}{M}(\sigma_a - \fl{\sigma_a})
\end{eqnarray}
with probability 
\begin{eqnarray}
&&  \label{eq:perfAnalysPr2} 
  \frac{\sin^2 (\pi(\sigma_a - \fl{\sigma_a}))}{M^2 \sin^2
    (\frac{\pi}{M}(\sigma_a - \fl{\sigma_a}))}\,
\left(1+(1-\delta_{\fl{\sigma_a},0})
\frac{\sin^2(\pi(\sigma_a-\fl{\sigma_a})/M)}
{\sin^2(\pi(\sigma_a+\fl{\sigma_a})/M)}\right) \nonumber \\
&\ge&   
\frac{\sin^2
    \big(\pi(\sigma_a - \fl{\sigma_a})\big)}{\pi^2(\sigma_a -
    \fl{\sigma_a})^2} =1- \frac{\pi^2}{3}(\sigma_a - \fl{\sigma_a})^2
  +O\big((\sigma_a - \fl{\sigma_a})^4\big) . 
\end{eqnarray}
\end{enumerate}
\end{PerfAnalys}
\vspace{24pt}

\begin{proof}
As before, let $\theta_a=\tha$ and 
$$
\br{S_M(\omega)}= 
\frac{1}{\sqrt{M}}
\sum_{k=0}^{M-1} e^{2\pi i \omega k} \br{k}, \quad i=\sqrt{-1}, 
$$ 
for arbitrary $\omega \in \R$. Note that
$$
F_{M,m}\br{j}=
\begin{cases}
\br{S_M(j/M)}& \text{for $j=0,1,\ldots,M-1$},\\
\br{j}& \text{for $j=M,M+1,\ldots,2^m-1$}.
\end{cases}
$$ 
The steps 1--3 of the \textbf{QS} algorithm are
equivalent to the application of the operator $(F_{M,m}^{-1} \otimes I
)\,\Lambda_m(Q_f)\, F_{M,m} \otimes W_N$ to the state $\br{0}\br{0}\in
\mathcal{H}_m \otimes \mathcal{H}_n$.  Then $\br{\eta_1}$ can be
written as \lb $M^{-1/2}\sum_{j=0}^{M-1}\br{j} \br{\psi}$, and 
$\br{\psi}=W_N\br{0}$ is given by \eqref{eq:represent}. Hence
\begin{align} 
\nonumber
 \br{\eta_1}= 
\frac{-i}{\sqrt{2M}}\sum_{j=0}^{M-1}\br{j} \left( e^{i
      \theta_a} \br{\psi_+}-e^{-i \theta_a} \br{\psi_-}\right).
\end{align}
Applying $\Lambda_m(Q_f)$ in Step 2 and remembering that $Q_f^j
\br{\psi_\pm}=\lambda^j_\pm \br{\psi_\pm}$, we obtain 
\begin{align*}
  \br{\eta_2}&=\Lambda_m(Q_f)\,\br{\eta_1}=
      \frac{-i}{\sqrt{2M}}\sum_{j=0}^{M-1}\br{j} \left(e^{2ij\theta_a}
      e^{i \theta_a} \br{\psi_+}-e^{-2ij\theta_a} e^{-i \theta_a}
      \br{\psi_-}\right) \\ &=\frac{-i}{\sqrt{2}} \left( e^{i
      \theta_a} \br{S_M(\sigma_a/M)}\br{\psi_+} - e^{-i \theta_a}
      \br{S_M(-\sigma_a/M)}\br{\psi_-}\right).
\end{align*}

Since $j=0,1,\dots,M-1$, the largest power of $Q_f$ is $M-1$.
Hence, we use exactly $M-1$ quantum queries to compute
$\br{\eta_2}$. The remaining steps of the \textbf{QS} algorithm 
do not use quantum queries. This means that the total number
of quantum queries used by the \text{QS} algorithm is $M-1$,
and obviously we are using $n+m$ qubits. 
This proves the first part of Theorem \ref{thm:PerfAnalys}.

Step 3 yields the state 
\begin{align*} 
\br{\eta_3}\,=\,(F_{M,m}^{-1}\otimes I) \br{\eta_2}
= \frac{-i}{\sqrt{2}} \left( e^{i \theta_a} F_{M,m}^{-1}
    \br{S_M(\sigma_a/M)} \br{\psi_+} - e^{-i\theta_a} F_{M,m}^{-1}
    \br{S_M(-\sigma_a/M)} \br{\psi_-}\right).
\end{align*}
We are ready to analyze the probability of the outcome $j$ of the \textbf{QS}
algorithm. Observe that
\begin{align*}
\br{\alpha_{\pm}}&:=\left(\br{j}\kt{j}\otimes I\right)\,
F^{-1}_{M,m}\br{S_M(\pm \sigma_a/M)}\br{\psi_{\pm}}\\
&=\langle j |F^{-1}_{M,m}\br{S_M(\pm \sigma_a/M)}\,\br{j}\br{\psi_{\pm}} \\
&= 
\begin{cases}
\inn{S_M(j/M)}{S_M(\pm \sigma_a/M)}\,\br{j}\br{\psi_{\pm}} & \text{for $j=0,1,\ldots,M-1$},\\
0 & \text{for $j=M,M+1,\ldots,2^m-1$},
\end{cases}
\end{align*}
and therefore 
\begin{equation*}
\inn{\alpha_{\pm}}{\alpha_{\pm}}=
\begin{cases}
\left|\langle S_M(j/M) \br{S_M(\pm \sigma_a/M)}\right|^2 
\langle \psi_{\pm}\br{\psi_{\pm}}
& \text{for $j=0,1,\ldots,M-1$}. \\
0 & \text{for $j=M,M+1,\ldots,2^m-1$}. 
\end{cases}
\end{equation*}
Observe that for $a\in(0,1)$, we have $\langle \psi_{\pm}\br{\psi_{\pm}}=1$,
whereas for $a\in\{0,1\}$, we have $\langle \psi_{+}\br{\psi_{+}}=~2$
and $\langle \psi_{-}\br{\psi_{-}}=0$. 

For $\omega_1, \omega_2 \in \R$ we have
\begin{align}
  \nonumber
  |\langle S_M(\omega_1)\br{S_M(\omega_2)}|^2 &=\left|
    \left(\frac{1}{\sqrt{M}} \sum_{j=0}^{M-1} e^{-2\pi i \omega_1 j}
      \kt{j}\right) \left(\frac{1}{\sqrt{M}} \sum_{j=0}^{M-1} e^{2\pi
        i \omega_2 j}
      \br{j}\right) \right|^2\\
\nonumber
  &=\frac{1}{M^2} \left| \sum_{j=0}^{M-1} e^{-2\pi i (\omega_1 -
      \omega_2)j} \right|^2.
\end{align}
If $\omega_1-\omega_2$ is an integer then the last sum is clearly $M$, and the
whole expression is $1$. If $\omega_1-\omega_2$ is not an integer then 
$$
\frac1M\,\sum_{j=0}^{M-1}e^{-2\pi i (\omega_1-\omega_2)j}\,=\,
\frac{e^{-2\pi iM(\omega_1-\omega_2)}-1}{M(e^{-2\pi i(\omega_1-\omega_2)}
-1)}, 
$$
which holds for all $\omega_1,\omega_2\in\R$ if we take $0/0$ as $1$.
Therefore
$$ 
\left|\frac1M\,\sum_{j=0}^{M-1}e^{-2\pi i
(\omega_1-\omega_2)j}\right|^2\,=\, \frac{1-\cos(2\pi
M(\omega_1-\omega_2))}{M^2(1-\cos(2\pi(\omega_1-\omega_2)))}\,=\,
\frac{\sin^2(\pi
M(\omega_1-\omega_2))}{M^2(\sin^2(\pi(\omega_1-\omega_2)))}.
$$
Hence
\begin{equation}\label{157}
   |\langle S_M(\omega_1)\br{S_M(\omega_2)}|^2 =
  \frac{\sin^2(M\pi(\omega_1-\omega_2))}{M^2 \sin^2(\pi
    (\omega_1- \omega_2))}
\end{equation}
which holds for all $\omega_1,\omega_2\in\R$ if we take $0/0$ as $1$.
Applying this we conclude that
$$
\langle \alpha_{\pm} \br{\alpha_{\pm}}= 
\begin{cases}
\frac{\sin^2(\pi(j\mp\sigma_a))}
{M^2\sin^2(\pi(j\mp\sigma_a)/M)}\langle \psi_{\pm}\br{\psi_{\pm}}
 & \text{for $j=0,1,\ldots,M-1$}, \\
0 & \text{for $j=M,M+1,\ldots,2^m-1$}. 
\end{cases}
$$

The outcome $j$ occurs after the measurement and the state $\br{\eta_3}$ 
collapses to the state
$(\langle \eta_3 | M_j^{\dag} M_j \br{\eta_3})^{-1} M_j\br{\eta_3}$, where
$M_j=\br{j}\kt{j}\otimes I$. For $j=0,1,\ldots, M-1$, we have 
\begin{multline*}
M_j\br{\eta_3}=\br{j}\bigg(\frac{-i}{\sqrt{2}}\big(
e^{i\theta_a}\langle S_M(j/M) \br{S_M(\sigma_a/M)}\,\br{\psi_{+}}
 \\ 
 - e^{-i\theta_a}\langle S_M(j/M) \br{S_M(-\sigma_a/M)}\,\br{\psi_{-}}
\big)\bigg),
\end{multline*}
whereas $M_j \br{\eta_3}=0$ for $j=M,M+1, \ldots,2^m-1$.
Since $\br{\psi_{+}}$ and $\br{\psi_{-}}$ are orthogonal we have
$$
\langle \eta_3 | M_j^{\dag}M_j\br{\eta_3}\,=\,
\tfrac12\left(
\langle \alpha_{+}\br{\alpha_{+}}\,+\,\langle \alpha_{-}\br{\alpha_{-}}\right).
$$
Hence, the outcome $j$, $j=0,1,\ldots,M-1$, occurs with probability
\begin{equation}\label{158}
p_f(j)\,=\,
\frac12\left(\frac{\sin^2(\pi(j-\sigma_a))}{M^2\sin^2(\pi(j-\sigma_a)/M)}\,+
\,\frac{\sin^2(\pi(j+\sigma_a))}{M^2\sin^2(\pi(j+\sigma_a)/M)}\right).
\end{equation}
Indeed, for $a\in(0,1)$, we have $\langle\psi_{\pm}\br{\psi_{\pm}}=1$
and (\ref{158}) follows from the form of 
$\langle \alpha_{\pm}\br{\alpha_{\pm}}$. For $a\in\{0,1\}$, we have
$\langle \psi_{+}\br{\psi_{+}}=2$ and $\langle \psi_{-}\br{\psi_{-}}=0$.
Since the two terms in (\ref{158}) are now the same, the formula for
$\langle \alpha_{+}\br{\alpha_{+}}$ again yields (\ref{158}). 

Since $\sin^2(\pi(j-\sigma_a))=\sin^2(\pi(j+\sigma_a))$, the last formula
is equivalent to  (\ref{eq:pj}). Obviously for $j=M,M+1,\ldots, 2^m-1$, 
the probability of the outcome $j$ is zero. This proves the second part
of Theorem~\ref{thm:PerfAnalys}.

Assume now that $\sigma_a \in \Z$. 
If $\sigma_a=0$ or $\sigma_a=\tfrac{1}{2}M$ (if $M$ is even) 
then the probability
$p_f(\sigma_a)$ of the outcome $\sigma_a$ is $1$.  For $\sigma_a=0$ we have
$a=0$ and the output is $\bar a_f(0)=0$. For $\sigma_a=\tfrac{1}{2}M$ we have 
$a=1$ and the output is $\bar a_f(\tfrac{1}{2}M)=1$. Hence, in both cases 
the \textbf{QS} algorithm outputs the exact value with probability $1$. 

If $\sigma_a\in \Z$ and $\sigma_a\notin\{0,\tfrac{1}{2}M\}$ 
then the probability of
the distinct outcomes $\sigma_a$ and $M-\sigma_a$ is $\tfrac{1}{2}$. 
These two values
of the outcomes yield the same output
$$
\sin^2\left(\pi\sigma_a/M\right)\,=\,\sin^2\left(\pi(M-\sigma_a)/M\right)\,=
\,a.
$$
Hence, the \textbf{QS} algorithm outputs the exact value with probability $1$.
This proves the third part of Theorem~\ref{thm:PerfAnalys}.

We now turn to the case when $\sigma_a \notin \Z$.
It is easy to check that the third part of Theorem~\ref{thm:PerfAnalys}
holds for $M=1$. Assume then that $M\ge2$ which implies that
$\cl{\tfrac{1}{2}M}\le M-1$. Since $\sigma_a$ is not an integer,
we have $\cl{\sigma_a}\ge1$, 
$\cl{\sigma_a}\le\cl{\tfrac{1}{2}M}\le M-1$ and $M-\cl{\sigma_a}\le M-1$. 
This means that both $\cl{\sigma_a}$ and $M-\cl{\sigma_a}$ may 
be the outcomes of the \textbf{QS} algorithm. 
Obviously, these two outcomes are different iff $\cl{\sigma_a}\not=
\tfrac{1}{2}M$. 
Similarly, both $\fl{\sigma_a}$ and $(1-\delta_{\fl{\sigma_a},0})M-
\fl{\sigma_a}$ may be also the outcomes. 
They are different iff $\fl{\sigma_a}\not=0$. 

We show that the outputs for the outcomes $\cl{\sigma_a}$ 
and $\fl{\sigma_a}$ satisfy \eqref{eq:perfAnalysErr1} and
\eqref{eq:perfAnalysErr2} with probabilities 
\eqref{eq:perfAnalysPr1} and \eqref{eq:perfAnalysPr2}, respectively.
We focus on the output for the outcome $\cl{\sigma_a}$ and its 
probability. The proof for the outcome $\fl{\sigma_a}$ is similar.

We  estimate the error of the \textbf{QS} algorithm for the
output $\bar a=\sin^2(\pi\cl{\sigma_a}/M)$. 
Recall that $\x=\pi(\cl{\sigma_a}-\sigma_a)/M$.
We have
\begin{align*}
  |\bar a- a| &= |\sin^2(\pi\cl{\sigma_a}/M)- 
    \sin^2(\pi\sigma_a/M)| =|\sin(\x)\, 
   \sin(\x+2\pi\sigma_a/M)|\\
   & =\big|\sin(\x)\,\big(\sin(2\pi\sigma_a/M)\cos(\x)+
\cos(2\pi \sigma_a/M)\sin(\x)\big)\big|    \\
   &\le \pi(\cl{\sigma_a}-\sigma_a)/M.
\end{align*}
Since $\sin(2\pi\sigma_a/M)=2\sqrt{a(1-a)}$ and $\cos(2\pi\sigma_a/M)=
1-2a$, this proves the estimate of the error of the \textbf{QS}
algorithm in the fourth part  of Theorem \ref{thm:PerfAnalys}.

We find the probability of the output $\bar a$. 
Since $\sin^2(\pi t/M)$ is injective for $t\in[0,\tfrac{1}{2}M]$,
the output $\bar a$ occurs only for the outcomes $\cl{\sigma_a}$
and $M-\cl{\sigma_a}$. 
If $\cl{\sigma_a}=\tfrac{1}{2}M$ then these two
outcomes are the same and $\bar a$ occurs with probability
$p_f(\frac{1}{2}M)$. Due to (\ref{158})
$$
p_f\left(\tfrac{1}{2}M\right)\,=\,
\frac{\sin^2\left(\pi(\tfrac{1}{2}M- 
\sigma_{a_f})\right)}
{M^2\sin^2\left(\pi(\tfrac{1}{2}M-\sigma_{a_f})/M\right)}
$$
which agrees with the claim in Theorem \ref{thm:PerfAnalys}. 

If $\cl{\sigma_a}\not=\tfrac{1}{2}M$ then $\cl{\sigma_a}\not=M-\cl{\sigma_a}$
and $\bar a$ occurs for exactly two distinct outcomes. 
The probability of $\bar a$ is now equal to the sum of the
probabilities $p_f(\cl{\sigma_a})+p_f(M-\cl{\sigma_a})$ with $p_f$'s given by
(\ref{158}). Since both terms are equal, the probability of $\bar a$
is $2p_f(\cl{\sigma_a})$ which also agrees with the claim 
in Theorem \ref{thm:PerfAnalys}. Since
$\sin(\frac{\pi}{M}(\cl{\sigma_a} - \sigma_a))  \le
\frac{\pi}{M}(\cl{\sigma_a} - \sigma_a)$ we have 
$$
\frac{\sin^2 (\pi(\cl{\sigma_a} - \sigma_a))}{M^2 \sin^2
    (\frac{\pi}{M}(\cl{\sigma_a} - \sigma_a))} \ge   \frac{\sin^2
    (\pi\cl{\sigma_a} - \sigma_a)}{\pi^2(\cl{\sigma_a} - \sigma_a)^2}.
$$
We finish proving \eqref{eq:perfAnalysPr1} 
using the standard expansion of the sine.
This completes the proof. 
\end{proof}
\vspace{12pt}

Based on Theorem \ref{thm:PerfAnalys} we 
present simplified estimates of the
error of the \textbf{QS} algorithm and of the corresponding
probability. 

\vspace{12pt}

\begin{MaxCor}
\label{MaxCor}
The \textbf{QS} algorithm  outputs $\bar a$ such that
\begin{equation}
  \label{eq:maxCorErr}
  | \bar a - a| \le \frac{\pi}{M} \max\{\cl{\sigma_a} - \sigma_a,
  \sigma_a - \fl{\sigma_a} \}
\end{equation}
with probability at least 
\begin{equation}
  \label{eq:maxCorPr}
   \frac{\sin^2 (\pi(\cl{\sigma_a} - \sigma_a))}{M^2 \sin^2
    (\frac{\pi}{M}(\cl{\sigma_a} - \sigma_a))}+  \frac{\sin^2
    (\pi(\sigma_a - \fl{\sigma_a}))}{M^2 \sin^2(\frac{\pi}{M}(\sigma_a
    - \fl{\sigma_a}))} \ge \frac{8}{\pi^2}.
\end{equation}
\end{MaxCor}
\begin{proof}
  It is enough to prove Corollary \ref{MaxCor} if $\sigma_a$ is not an
  integer. 
  The estimate of the error of the \textbf{QS}
  algorithm by the
  maximum of the estimates \eqref{eq:perfAnalysErr1} and
  \eqref{eq:perfAnalysErr2} holds with probability that is the sum
  of the probabilities \eqref{eq:perfAnalysPr1}
  and~\eqref{eq:perfAnalysPr2}. Moreover, $\cl{\sigma_a}-\sigma_a = 1-
  (\sigma_a - \fl{\sigma_a})$. It now suffices to observe that the
  function
\begin{equation*}
  g(\Delta)=\frac{\sin^2(\pi\Delta)}{
  \pi^2 \Delta^2}+\frac{\sin^2(\pi(1-\Delta))}{
  \pi^2 (1-\Delta)^2}
\end{equation*}
is a lower bound of the left hand side of \eqref{eq:maxCorPr} with
$\Delta = \cl{\sigma_a} - \sigma_a$, and
attains the minimum $\tfrac{8}{\pi^2}$ on the interval $[0, 1]$ for
$\Delta=\tfrac{1}{2}$, see also \cite{BMHT}.
\end{proof}
\vspace{12pt}

Corollary \ref{MaxCor} guarantees high probability of the estimate
\eqref{eq:maxCorErr}. Unfortunately this estimate does not preserve
the continuity of the estimates \eqref{eq:perfAnalysErr1} and
\eqref{eq:perfAnalysErr2} with respect to $\cl{\sigma_a} - \sigma_a$
and $\sigma_a - \fl{\sigma_a}$. The continuity of the estimates will
be present in the next corollary at the expense of the probability of the
outcome. This corollary will also play an essential role in the study
of the average-probabilistic  error of the \textbf{QS} algorithm.

\vspace{12pt}
\begin{MinCor}
\label{MinCor}
The \textbf{QS} algorithm   outputs $\bar a$ such that 
\begin{equation}
  \label{eq:minCorErr}
  | \bar a - a| \le \frac{\pi}{M} \min\{\cl{\sigma_a} - \sigma_a,
  \sigma_a - \fl{\sigma_a} \}
\end{equation}
with probability at least 
\begin{equation}
  \label{eq:minCorPr}
   \max \left\{\frac{\sin^2 (\pi(\cl{\sigma_a} - \sigma_a))}{M^2 \sin^2
    (\frac{\pi}{M}(\cl{\sigma_a} - \sigma_a))},  \frac{\sin^2
    (\pi(\sigma_a - \fl{\sigma_a}))}{M^2 \sin^2(\frac{\pi}{M}(\sigma_a
    - \fl{\sigma_a}))}\right\} \ge \frac{4}{\pi^2}.
\end{equation}
\end{MinCor}
\begin{proof}
We may again assume that $\sigma_a$ is not an integer. Let us define
\begin{equation*}
  w(\Delta)=\frac{\sin^2(\pi\Delta)}{M^2 \sin^2(\frac{\pi}{M}
    \Delta)}\qquad \mathrm{for}\quad\Delta\in [0,1].
\end{equation*}

Then $w(\cl{\sigma_a}-\sigma_a)$ is the probability
of~\eqref{eq:perfAnalysErr1} and $w(1-(\cl{\sigma_a}-\sigma_a))$ is
the probability of~\eqref{eq:perfAnalysErr2}. 
For $\Delta\in[0,\tfrac{1}{2}]$, note that $w(\cdot)$ is
decreasing, and $w(1-\cdot)$ is increasing. Therefore
\begin{equation*}
  w(\Delta) \ge w(\tfrac{1}{2})\ge w(1-\Delta) \qquad \mathrm{for}\quad
  \Delta\in[0,\tfrac{1}{2}]. 
\end{equation*}

Let $\cl{ \sigma_a} -\sigma_a \le \sigma_a - \fl{\sigma_a}$. Then
$\cl{\sigma_a} - \sigma_a \le \tfrac{1}{2}$. In this case 
\eqref{eq:minCorErr} is
equivalent to \eqref{eq:perfAnalysErr1} and holds with probability
at least $w( \cl{ \sigma_a} -\sigma_a)$, which corresponds to
\eqref{eq:minCorPr}. Analogously, if $\cl{ \sigma_a} -\sigma_a \ge
\sigma_a - \fl{\sigma_a}$ then $\cl{\sigma_a} - \sigma_a \ge \tfrac{1}{2}$.
In this case \eqref{eq:minCorErr} is equivalent to 
\eqref{eq:perfAnalysErr2} and
holds with probability at least $w(\sigma_a- \fl{\sigma_a})$, which also
corresponds to \eqref{eq:minCorPr}. Finally, note that 
\begin{equation*}
  \max \left\{ w(\cl{ \sigma_a} -\sigma_a), w(\sigma_a -
\fl{\sigma_a})\right\}
\end{equation*}
is minimal for $\cl{ \sigma_a} -\sigma_a= \tfrac{1}{2}$ and is equal to
\begin{equation*}
  \frac{1}{M^2} \sin^{-2} \frac{\pi}{2M} \ge \frac{4}{\pi^2}.
\end{equation*}
\end{proof}
\vspace{12pt}

Unfortunately for $\cl{\sigma_a}-\sigma_a$ close to $\tfrac{1}{2}$ the
probability of the estimate \eqref{eq:minCorErr} is too small.
However, in this case we may use Corollary \ref{MaxCor}, which yields
the estimate with high probability.

We now turn to global error estimates, that is, estimates independent of
$a$.  Theorem~\ref{thm:Brass&co} of \cite{BMHT} states, in particular,
that $|\bar a - a| \le \pi/M + \pi^2/M^2$
with probability at least $\tfrac{8}{\pi^2}$.  We now improve this estimate
by combining the estimates \eqref{eq:maxCorErr} and
\eqref{eq:minCorErr}.

\begin{ImprovedCor}
\label{ImprovedCor}
The \textbf{QS} algorithm   outputs $\bar a$ such that 
\begin{equation} 
  \label{eq:improvedCorErr}
  | \bar a - a| \le \frac{3}{4}\,\frac{\pi}{M}
\end{equation}
with probability at least $\tfrac{8}{\pi^2}$. That is, 
$$
e^{\wpp}(M,\tfrac{8}{\pi^2})\,\le\,\frac34\,\frac{\pi}M.
$$

\end{ImprovedCor}
\begin{proof}
Let us define
\begin{equation}
\label{eq:hDelta}
  h(\Delta)=\max \left\{ \frac{\sin^2(\pi\Delta)}{
  \pi^2 \Delta^2},\frac{\sin^2(\pi(1-\Delta))}{
  \pi^2 (1-\Delta)^2} \right\}.
\end{equation}
Clearly, $h(\cl{\sigma_a}- \sigma_a)$ is a lower bound of the
$\max\{w(\cl{\sigma_a}- \sigma_a), w(1-(\cl{\sigma_a}- \sigma_a))\}$
and therefore $h(\cl{\sigma_a}-\sigma_a)$ is a lower bound of the
probability of the output satisfying (\ref{eq:minCorErr}).  We consider
two cases.

Assume first that $\Delta=\cl{\sigma_a} -\sigma_a \in [0, \tfrac{1}{4}] \cup
[\tfrac{3}{4},1]$. It is easy to see that then \lb $h(\Delta) \ge 
\tfrac{8}{\pi^2}$ and
the estimate~\eqref{eq:minCorErr} yields
$$
|\bar a -a|\le \frac{\pi}{M} \min\{\cl{\sigma_a} - \sigma_a,
  \sigma_a - \fl{\sigma_a} \}\le \frac{1}{4}\;\frac{\pi}{M}
$$
with probability at least $\tfrac{8}{\pi^2}$. 

Assume now that $\cl{\sigma_a}- \sigma_a \in (\tfrac{1}{4},
\tfrac{3}{4})$. Then we can
use the estimate~\eqref{eq:maxCorErr}, which holds unconditionally with
probability at least $\tfrac{8}{\pi^2}$. In this case, we have
$$
|\bar a -a|\le \frac{\pi}{M}\max\{\cl{\sigma_a} - \sigma_a,
  \sigma_a - \fl{\sigma_a} \} \le \frac{3}{4}\;\frac{\pi}{M}.
$$
These two estimates combined together yield \eqref{eq:improvedCorErr}.
\end{proof}
\vspace{12pt} The obvious consequence of Corollary \ref{ImprovedCor}
is that for $M$ large enough we can compute the value of $a$ exactly
by rounding the output.
\begin{IntegerCor}
Assume that 
\begin{equation*}
  M > \frac{3\pi}{2}N.
\end{equation*}
Then the rounding of the \textbf{QS} algorithm output to the nearest
number of the form $k/N$ yields the exact value of the sum $a$ with
probability at least $\tfrac{8}{\pi^2}$.
\end{IntegerCor}
\vspace{12pt}

The proof of Corollary \ref{ImprovedCor} may suggest that the constant
$\tfrac{3}{4}$ in \eqref{eq:improvedCorErr} can be decreased. Furthermore one
may want to decrease the constant $\tfrac{3}{4}$ at the expense of
decreasing the probability $\tfrac{8}{\pi^2}$. These points are addressed in
the next corollary. We shall see that the constant $\tfrac{3}{4}$ may be
lowered only by decreasing the probability.

\vspace{12pt}
\begin{GlobalCor}
\label{GlobalCor}
Define
\begin{equation*}
  C(p)=\inf \bigg\{C:\;|\bar a_f -a_f| \le C\frac{\pi}{M}\ \ 
\forall f\in\B_N\ \text{with
      probability at least}\; p \bigg\}
\end{equation*}
and
\begin{equation*}
  v(\Delta)=\frac{\sin^2 (\pi \Delta) }{\pi^2 \Delta^2}\qquad
  \text{for}\ \; \Delta\in [\tfrac{1}{4},\tfrac{1}{2}].
\end{equation*}
Then
\begin{equation}
  \label{eq:cpEstimate}
  C(p) \le \begin{cases}
\frac{1}{2} &\text{for}\quad p \in [0, \tfrac{4}{\pi^2}),\\
1-v^{-1}(p) &\text{for}\quad p \in [\tfrac{4}{\pi^2}, \tfrac{8}{\pi^2}], \\
M/\pi &\text{for} \quad p \in (8/ \pi^2, 1]. \\
\end{cases}
\end{equation}
Moreover, $1-v^{-1}(p) \in [\tfrac{1}{2},\tfrac{3}{4}]$ and 
\begin{equation}
  \label{eq:linearEstim}
  \left|\frac{\pi^2}{16}p+\frac14 -\left(1-v^{-1}(p)\right)\right|
\,\le\, 0.0085 \quad \mbox{for}\ \ p\in[\tfrac{4}{\pi^2}
,\tfrac{8}{\pi^2}].
\end{equation}
\end{GlobalCor}
\begin{proof}
  For $p\in [0, \tfrac{4}{\pi^2})$, Corollary \ref{GlobalCor} is a consequence
  of Corollary \ref{MinCor}.  For $p\in (\tfrac{8}{\pi^2}, 1]$, 
  Corollary~\ref{GlobalCor} trivially holds since $| \bar a - a| \le 1
  = (M/\pi)
  \pi/M$.  For the remaining $p$'s we use a proof technique similar
  to that of Corollary \ref{ImprovedCor}.
  
  Let $p\in [\tfrac{4}{\pi^2}, \tfrac{8}{\pi^2}]$. 
  It is easy to check that $v$ is
  decreasing and, therefore, $v^{-1}(p)$ is well defined and
  $v^{-1}(p) \in [\tfrac{1}{4}, \tfrac{1}{2}]$. 
We have to show that the estimate
\begin{equation}
  \label{eq:v}
  |\bar a - a| \le (1- v^{-1}(p))\frac{\pi}{M} 
\end{equation}
holds with probability at least $p$.  We consider two cases.

Assume first that $\Delta=\cl{\sigma_a} -\sigma_a \in [0, v^{-1}(p)]
\cup [1-v^{-1}(p),1]$. Observe that the function~$h$ defined in
\eqref{eq:hDelta} can be rewritten as
\begin{equation*}
  h(\Delta)=\max\{v(\Delta), v(1-\Delta)\}.
\end{equation*}
It is easy to see that in this case, $h(\Delta) \ge p$ and the
estimate \eqref{eq:minCorErr} yields
$$
|\bar a -a|\le \frac{\pi}{M} \min\{\Delta, 1-\Delta\}\le
v^{-1}(p)\;\frac{\pi}{M}\leq(1-v^{-1}(p)) \frac{\pi}{M}
$$  with probability at least $p$. 

Assume now that $\Delta=\cl{\sigma_a}- \sigma_a \in
(v^{-1}(p),1-v^{-1}(p))$. Then we can use the
estimate~\eqref{eq:maxCorErr}, which holds unconditionally with 
probability at least $\tfrac{8}{\pi^2}>p$. In this case, we have
$$
|\bar a -a|\le \frac{\pi}{M}\max\{\Delta,
  1-\Delta \} \le (1-v^{-1}(p))\frac{\pi}{M}.
$$
This proves \eqref{eq:v}.

We found the estimate~\eqref{eq:linearEstim} by numerical
computations. 
\end{proof}
\vspace{24pt}

\newpage

\begin{figure}[!ht]
\begin{center}
\includegraphics[scale=0.5]{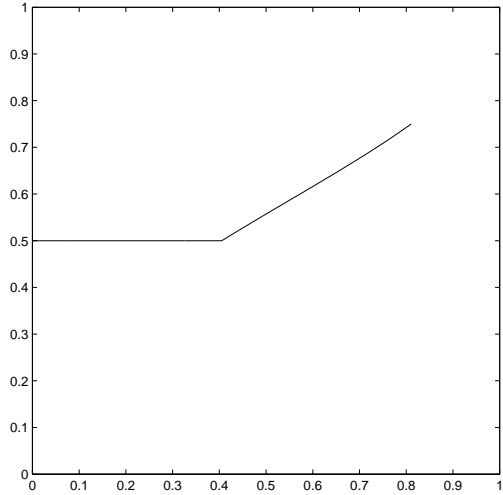}
\caption{The estimate \eqref{eq:cpEstimate} of $C(p)$ for $p\in[0,
\tfrac{8}{\pi^2}]$}
\label{fig:cp}
\end{center}
\end{figure}

{}From Figure \ref{fig:cp} we see that the estimate
\eqref{eq:cpEstimate} is almost linear on the interval $[\tfrac{4}{\pi^2},
\tfrac{8}{\pi^2}]$, which explains why the right hand side of
the estimate \eqref{eq:linearEstim} is small.

\begin{figure}[!ht]
\begin{center}
\includegraphics[scale=0.5]{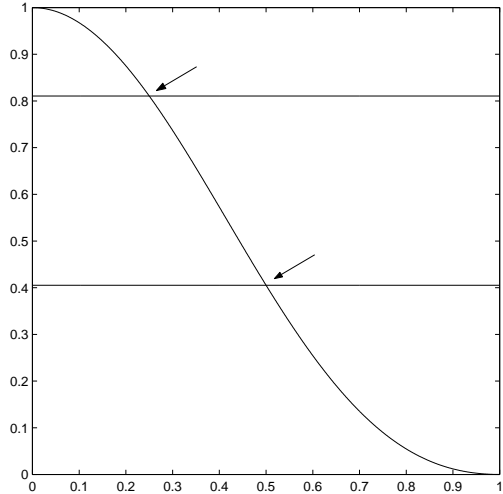}
\caption{
The function $v$ on $[0,1]$. The two horizontal
  lines show $\tfrac{4}{\pi^2}$ and $\tfrac{8}{\pi^2}$ levels. 
The part of the graph
  between the arrows shows that $v$ is almost linear.}
\label{fig:v}
\end{center}
\end{figure}
 
\vskip 2pc
We now find a sharp bound on the worst-probabilistic error of the
\textbf{QS} algorithm. We show that for large $M$ and $N/M$
the bound obtained in Corollary \ref{GlobalCor} is optimal
for $p\in(\tfrac{1}{2},\tfrac{8}{\pi^2}]$.

\begin{asymppp} 
\label{thm:asympp}
For large $M$ and $N/M$, the worst-probabilistic error of the 
\textbf{QS} algorithm is given by
$$
e^{\wpp}(M,p)\,=\,(1-v^{-1}(p))\frac{\pi}M\,(1+O(M^{-1})+O(MN^{-1})\,)\qquad
\mbox{for}\ p\in(\tfrac{1}{2},\tfrac{8}{\pi^2}].
$$
Here, $v$ is as in Corollary \ref{GlobalCor}, and
$1-v^{-1}(p) \approx (\pi^2/16)p+\tfrac{1}{4}$ by (\ref{eq:linearEstim}).  
\end{asymppp}
\begin{proof}
{}From Corollary \ref{GlobalCor}, it is enough to show a lower bound
on the error. Define
$$
s_1=\sin^2\left(\frac{\pi\cl{\tfrac{1}{4}M}}{M}\right)\quad\mbox{and}\quad 
s_2=\sin^2\left(\frac{\pi(1+\cl{\tfrac{1}{4}M})}{M}\right).
$$
For large $M$, we have
$$
s_i=\tfrac{1}{2}+O(M^{-1})\quad\mbox{and}\quad s_2-s_1=
\sin\left(\frac{\pi}{M}\right)
\sin\left(\frac{\pi(1+2\cl{\tfrac{1}{4}M})}{M}\right)=\frac{
(1+O(M^{-1}))\pi}{M}.
$$
There exist two Boolean 
functions $f_1$ and $f_2$ with sums $a_1=a_{f_1}$ and $a_2=a_{f_2}$
such that 
$$
|a_i-s_i|\le N^{-1}\quad\mbox{for}\quad i=1,2.
$$
Since $\sigma_{s_i}=\cl{\tfrac{1}{4}M} +(i-1)$ and the derivative of $\sigma_a$
for $a=\tfrac{1}{2}$ is $M/\pi$, we have
$$
\sigma_{a_1}=\cl{\tfrac{1}{4}M}+O(MN^{-1})\quad\mbox{and}\quad
\sigma_{a_2}=\cl{\tfrac{1}{4}M}+1+ O(MN^{-1}).
$$
Obviously, $a_i=k_i/N$ for some integers $k_i$ with $k_1<k_2$.
Consider $\sigma_{x/N}$ for $x\in\{k_1,k_1+1,\dots,k_2\}$. Then
$\sigma_{x/N}$ varies from $\sigma_{a_1}$ for $x=k_1$ to $\sigma_{a_2}$
for $x=k_2$. 
Since $v^{-1}(p)\in[\tfrac{1}{4},\tfrac{1}{2})$, 
for a positive and small $\eta$ (we finally let $\eta$ go to
zero), we can choose $x=x_\eta$ such that for $a^*=x_\eta/N$ we have
$$
\sigma_{a^*}\,:=\,\cl{\tfrac{1}{4}M}+v^{-1}(p)+\eta +O(MN^{-1}).
$$
For large $N/M$, we then have
\begin{eqnarray*}
\fl{\sigma_{a^*}}=\cl{\tfrac{1}{4}M}\quad&\mbox{and}&\quad\cl{\sigma_{a^*}}=
\cl{\tfrac{1}{4}M}+1,\\
\sigma_{a^*}-\fl{\sigma_{a^*}}=v^{-1}(p)+\eta+O(M/N)\quad&\mbox{and}&\quad
\cl{\sigma_{a^*}}-\sigma_{a^*}=1-v^{-1}(p)-\eta+O(M/N).
\end{eqnarray*}
Let $\bar a^*_1$ denote the output for the outcome $\cl{\sigma_{a^*}}$, and
$\bar a^*_2$ for $\fl{\sigma_{a^*}}$. 

Due to (\ref{eq:perfAnalysErr1}) and (\ref{eq:perfAnalysErr2}) of 
Theorem \ref{thm:PerfAnalys} we have 
\begin{eqnarray*}
|a^*-\bar a^*_1|\,&=&\,
\frac{\pi}M\left(1-v^{-1}(p)-\eta\right)\left(1+O(M^{-1}+MN^{-1})\right)\\
|a^*-\bar a^*_2|\,&=&\,
\frac{\pi}M\left(v^{-1}(p)+\eta\right)\left(1+O(M^{-1}+MN^{-1})\right).
\end{eqnarray*}
Let us write $1+o(1)$ for $1+O(\eta^2+M^{-1}+MN^{-1})$. 
The probability of $\bar a^*_2$ is given by (\ref{eq:perfAnalysPr2}) and is
now equal to
$$
\frac{\sin^2(\pi(v^{-1}(p)+\eta))}{(\pi(v^{-1}(p)+\eta))^2}\,
\left(1+o(1)\right)\,=\,
\frac{\sin^2(\pi v^{-1}(p))+\pi\eta\sin(2\pi v^{-1}(p))}
{\pi^2v^{-1}(p)^2(1+2\eta/v^{-1}(p))}
\,\left(1+o(1)\right).
$$
Since $p=v(v^{-1}(p))=\sin^2(\pi v^{-1}(p))/(\pi v^{-1}(p))^2$, the probability
of $\bar a_2^*$ is
$$
p\,\left(1-2\eta\left(\frac1{v^{-1}(p)}-\pi\cot(\pi v^{-1}(p))\right)\right)
\,\left(1+o(1)\right).
$$
Since $\cot(t)< 1/t$ for $t\in[\tfrac{1}{4}\pi,\tfrac{1}{2}\pi]$, 
we see that the probability
of $\bar a_2^*$ is slightly less than~$p$ for small $\eta$. 

We are ready to find a lower bound on the worst-probabilistic error
$$
e^{\wpp}(M,p)=\max_{f \in \B_N} \;\min_{A:\; \mu(A,f) 
\ge p}\; \max_{j\in A} |a_f - \bar a_f(j)|
$$
of the \textbf{QS} algorithm.
Take the function $f$ that corresponds to $a^*$. We claimed that the error
is minimized if $A=\{\fl{\sigma_{a^*}},\cl{\sigma_{a^*}}\}$. Indeed,
$\fl{\sigma_{a^*}}$ must belong to $A$ since otherwise $\mu(A,f)\le
1-p_f(\fl{\sigma_{a^*}})=1-p+o(1)< p$ for $p>\tfrac{1}{2}$. The probability
of $\fl{\sigma_{a^*}}$ is slightly less than $p$, and so 
the set $A$ must also contain some other outcome $j$. If $j=\cl{\sigma_{a^*}}$
then the error bound is roughly $(1-v^{-1}(p)-\eta)\pi/M$, and the sum of the
probabilities of the outputs for the outcome $\fl{\sigma_{a^*}}$ and
$\cl{\sigma_{a^*}}$ is always at least $\tfrac{8}{\pi^2}\ge p$. 
On the other hand, if 
$\cl{\sigma_{a^*}}$ does not belong to the set $A$ then any other outcome
$j$ yields the output $\sin^2(\pi j/M)$. Since
$\sin^2(\pi(j+1)/M)-\sin^2(\pi j/M)=\sin(\pi/M)\sin(\pi(2j+1)/M)$,
the distribution of the outcomes around $\tfrac{1}{2}$ 
is a mesh with step size
roughly $\pi/M$. Hence, if $j\not=\cl{\sigma_{a^*}}$, the error is at least
roughly $(1+v^{-1}(p))\pi/M>\pi(1-v^{-1}(p))/M$. 
Thus the choice $j=\cl{\sigma_{a^*}}$ minimizes the error
and for $\eta$ tending to zero, the error is roughly $(1-v^{-1}(p))\pi/M$.
This completes the proof.
\end{proof}

{}From these results,  it is obvious how to guarantee that the
error of the \textbf{QS} algorithm is at most $\ve$ with 
probability at least $p$. Since $|\bar a - a|\le (1-v^{-1}(p))\pi/M$
holds with probability $p$, it is enough to take 
$M \ge (1-v^{-1}(p))\pi/\ve$. Due to Theorem
\ref{thm:asympp} this bound is sharp for small $\ve$ and large $\ve N$.
We have 
\begin{EpsilonCor}
\label{EpsilonCor} 
For $p\in(\tfrac{1}{2},\tfrac{8}{\pi^2}]$, the algorithm 
\textbf{QS}($f$, $\cl{(1-v^{-1}(p))\pi/\ve}$)
computes $\bar a$ with the error $\ve$  and probability at least
$p$ with $\cl{(1-v^{-1}(p))\pi/\ve}-1$
quantum queries. For small $\ve$ and large $\ve N$, the estimate
of the number of quantum queries is sharp. 
\end{EpsilonCor}

\subsection{Average-Probabilistic Error}
\label{sec:ACP}

In this section we study the average performance of the \textbf{QS}
algorithm with respect to some measure on the set $\B_N$ of all Boolean
functions defined on the set $\{0,\ldots, N-1\}$.
We consider two such measures. The first measure $\pr_1$ is uniformly
distributed on the set~$\B_N$, i.e.,
\begin{equation*}
\pr_1(f)= 2^{-N} \qquad \forall f\in \B_N.  
\end{equation*}
The second measure $\pr_2$ is uniformly distributed on the set of
results, i.e.,
\begin{equation*}
  \pr_2(f)=\frac{1}{ {N\choose k}(N+1)}\quad\text{if}\quad a_f=\frac{k}{N}.
\end{equation*}
For the average-probabilistic error we want to estimate 
\begin{equation*}
 e^{\ap}_{\pr_i}(M,p)=\sum_{f \in \B_N} \pr_i(f) \min_{A:\;
   \mu(A,f) \ge p} \max_{j\in A} |a_f - a_f(j)|\quad
   \text{for}\;i=1,2.
\end{equation*}

For the measures $\pr_i$, the mean of the random variable $a_f$ 
is clearly $\tfrac{1}{2}$. However, their first (central) moments
are very different. As we shall see,
the moment for the measure $\pr_1$ is small since it is of order
$N^{-1/2}$ whereas the moment for measure $\pr_2$ is roughly $\tfrac{1}{4}$. 
Since the first moments are the same as the error of the constant algorithm 
$\bar a_f(j)=\tfrac{1}{2}$, we can achieve small error of order $N^{-1/2}$
for the measure $\pr_1$ without any quantum queries, while this property
is not true for the measure $\pr_2$.

We now consider the measure $\pr_1$. It is interesting to ask
if the \textbf{QS} algorithm  has the same property as the constant
algorithm. We shall prove that this is indeed the case
iff $M$ is divisible by 4. 

We compute the first moment or the error of 
the constant algorithm, which is 
$$
\sum_{k=0}^N 2^{-N} {N \choose k}
  \left|\frac{1}{2}-\frac{k}{N}\right|.
$$
We do it only for odd $N$
since the case of even $N$ is analogous. We have
\begin{align*}
\sum_{k=0}^N 2^{-N} {N \choose k} \left|\frac{1}{2}-\frac{k}{N}\right|
  &= 2 \sum_{k=0}^{\fl{N/2}} {N  \choose  k} 
\left( \frac{1}{2}- \frac{k}{N} \right) =\sum_{k=0}^{\fl{N/2}} {N
  \choose  k} -2 \sum_{k=0}^{\fl{N/2}-1} {N -1
  \choose  k} \\ & = 2^{N-1}- 2 \frac{1}{2} \left(2^{N-1}-{N-1 \choose
  (N-1)/2} \right) = {N-1 \choose (N-1)/2}.
\end{align*}
Thus  
\begin{equation}
\label{eq:zero_err}
  e_{\pr_1}^{\ap}(0,1)=
\sum_{k=0}^N 2^{-N} {N \choose k}
  \left|\frac{1}{2}-\frac{k}{N}\right|=
\begin{cases}
  2^{-N} {N-1 \choose (N-1)/2} & \text{if $N$ is odd}, \\
  2^{-(N+1)} {N \choose N/2} & \text{if $N$ is even}.
\end{cases}
\end{equation}
By Stirling's formula 
\begin{equation*}
k!=\sqrt{2\pi k}
\left(\frac{k}{e}\right)^k e^{\theta_k/12k}\qquad\text{for
  certain $\theta_k \in [0,1]$ },
\end{equation*}
we estimate the both binomial quantities in  (\ref{eq:zero_err}) by
$$
    \frac{1}{\sqrt{2\pi}}\,\frac{1}{\sqrt{N-1}}\;e^{1/(12(N-1))} 
$$
proving that 
\begin{equation}\label{needed}
e^{\ap}_{\pr_1} (0,1) = \sum_{k=0}^N 2^{-N} {N \choose k} \left|
\frac{1}{2} - \frac{k}{N} \right| = \frac{1}{\sqrt{2 \pi}}
\frac{1}{\sqrt{N}} (1+o(1)) \le \frac{1}{\sqrt{2 \pi}}
\frac{1}{\sqrt{N-1}}\, e^{1/(12(N-1))}.
\end{equation}

We are ready to analyze the average-probabilistic error of the \textbf{QS} 
algorithm. 

\begin{WA4}
\label{thm:WA4}
  Assume that $M$ is divisible by $4$ and let $p\in(\tfrac{1}{2},
\tfrac{8}{\pi^2}]$. 
Then the average-probabilistic
  error of the \textbf{QS} algorithms with respect to the measure
  $\pr_1$ satisfies
\begin{align*} 
  e^{\ap}_{\pr_1}(M, p) & \le 
\min \bigg\{\frac{3}{4}\, \frac{\pi}{M}, \sqrt{\frac{3}{2\pi}} 
  \sqrt{1+\frac{\pi^2}{4M^2}}\,\frac{1}{\sqrt{N-1}}
  e^{1/(12(N-1))}\bigg\} \\ &\le \frac{3}{4} \pi (1+o(1)) \min
    \bigg\{\frac{1}{M}, \frac{1}{\sqrt{N}} \bigg\}.
\end{align*}
\end{WA4}
\begin{proof}
The estimate $e^{\ap}_{\pr_1}(M, p) \le e^{\ap}_{\pr_1}(M, \tfrac{8}{\pi^2})
\le e^{\wpp}_{\pr_1}(M,\tfrac{8}{\pi^2})$ is obvious and, applying Corollary~\ref{ImprovedCor}, we
  get   
\begin{equation*}
    e^{\ap}_{\pr_1}(M, \tfrac{8}{\pi^2}) \le \frac{3}{4}\, \frac{\pi}{M}. 
\end{equation*}

  As before denote $\sigma_a=\frac{M}{\pi} \arcsin \sqrt{a}$. Let $a=
  \tfrac{1}{2}+x$. We are interested in the behavior of
  $\sigma_{\tfrac{1}{2}+x}$ for $|x| < \tfrac{1}{2}$. 
  Clearly $\sigma_{\tfrac{1}{2}}=\tfrac{1}{4}M$. 
  Let $|x|< \tfrac{1}{2}$, By Taylor's theorem, we have 
  \begin{equation*}
    \sigma_{\tfrac{1}{2}+x} = \frac{M}{4} + \frac{M}{\pi} \frac{x}{2
    \sqrt{(1-\xi_x) \xi_x}}\qquad \text{for} \quad \xi_x \in (
  \tfrac{1}{2}, \tfrac{1}{2}+x)
  \end{equation*}
and $2 \sqrt{(1-\xi_x)\xi_x} \geq \sqrt{1-4x^2}$. Assume additionally
that 
\begin{equation*}
  \frac{M}{\pi} \frac{|x|}{\sqrt{1-4x^2}} \le \frac{1}{4}
\end{equation*}
which is equivalent to assuming that 
\begin{equation*}
  |x| \le \frac{\pi}{(16M^2+4\pi^2)^{1/2}}.
\end{equation*}
Since $M$ is divisible by $4$ then $\fl{\sigma_{\tfrac{1}{2}+x}}=
\tfrac{1}{4}M$ for $x
\geq 0$, and $\cl{\sigma_{\tfrac{1}{2}+x}}=\tfrac{1}{4}M$ for $x \leq 0$.
This yields
\begin{equation}
\label{eq:min_estim}
  \min\bigg\{\cl{\sigma_{\tfrac{1}{2}+x}}- \sigma_{\tfrac{1}{2}+x},  
   \sigma_{\tfrac{1}{2}+x}-
  \fl{\sigma_{\tfrac{1}{2}+x}} \bigg\}  \le \frac{M}{\pi}
  \frac{|x|}{\sqrt{1-4x^2}}.
\end{equation}
Observe that  
\begin{equation*}
\cl{\sigma_{\tfrac{1}{2}+x}} - \sigma_{\tfrac{1}{2}+x}
\in [0,\tfrac{1}{4}] \cup [\tfrac{3}{4},1].
\end{equation*}
Indeed, for $x \le 0$ we have 
$$ 
\cl{\sigma_{\tfrac{1}{2}+x}} - \sigma_{\tfrac{1}{2}+x} = 
\frac{M}{\pi} \frac{|x|}{2
  \sqrt{(1-\xi_x) 
\xi_x}} \in [0, \tfrac{1}{4}],
$$ 
and for $x \ge 0$ we have
$$
\cl{\sigma_{\tfrac{1}{2}+x}} - \sigma_{\tfrac{1}{2}+x} = 1-\frac{M}{\pi} \frac{|x|}{2
  \sqrt{(1-\xi_x) \xi_x}} \in [1- \tfrac{1}{4}, 1]=[\tfrac{3}{4},1],
$$
as claimed.

Let $a=\tfrac{1}{2}+x$. By the proof of Corollary \ref{ImprovedCor}, the
error of the \textbf{QS} algorithm satisfies
$$
|\bar a - a| \le \frac{\pi}{M} \min \left\{ \cl{\sigma_{\tfrac{1}{2}+x}} -
\sigma_{\tfrac{1}{2}+x}, \sigma_{\tfrac{1}{2}+x} - \fl{\sigma_{\tfrac{1}{2}
+x}} \right\}
$$
and by (\ref{eq:min_estim}), we have  
$$
|\bar a - a| \le \frac{|x|}{\sqrt{1-4x^2}}. 
$$

We split the sum that defines $e^{\ap}_{\pr_1}(M,\tfrac{8}{\pi^2})$ into
two sums. The first sum is for $f \in \B_N$ for which $a= a_f
= \tfrac{1}{2} +x$ with $|x| \le \pi / (16 M^2 + 4 \pi^2)^{1/2}$ and the second
sum is for $f$ for which $a= a_f = \tfrac{1}{2} +x$ with $|x| > \pi / (16 M^2
+ 4 \pi^2)^{1/2}$. For the first sum we estimate the error of the
\textbf{QS} algorithm by $|x| / \sqrt{1-4 x^2}$ and for the second
sum, by the worst-case error $3 \pi/(4 M)$. Hence we have 
\begin{align*}
  e^{\ap}_{\pr_1}(M,\tfrac{8}{\pi^2}) &\le \sum_{f:\,|a_f-\tfrac{1}{2}| \le \pi/(16
    M^2 + 4 \pi^2)^{1/2}}\pr_1(f) \frac{|a_f -
    \tfrac{1}{2}|}{\sqrt{1-4(a_f-\tfrac{1}{2})^2}} \\ 
&+   \sum_{f:\,|a_f-\tfrac{1}{2}| > \pi/(16 M^2 + 4 \pi^2)^{1/2}} \pr_1(f)
    \frac{3}{4} \frac{\pi}{M}.
\end{align*}
Since $a_f= k/N$ for some integer $k \in [0, N]$, 
\begin{align*}
\label{eq:first_sum}
  e^{\ap}_{\pr_1}(M,\tfrac{8}{\pi^2}) &\le \sum_{k:\,|k/N-\tfrac{1}{2}| \le \pi/(16
    M^2 + 4 \pi^2)^{1/2}} 2^{-N} {N \choose k}
  \frac{|k/N-\tfrac{1}{2}|}{\sqrt{1-4(k/N-\tfrac{1}{2})^2}} \\
\nonumber
& +    \frac{3}{4} \frac{\pi}{M} \sum_{k:\,|k/N-\tfrac{1}{2}| >
    \pi/(16 M^2 + 4 \pi^2)^{1/2}} 2^{-N} {N \choose k}.
\end{align*}
Since $1-4(k/N- \tfrac{1}{2})^2 \ge 1- \pi^2/(4M^2) \ge \tfrac{3}{4}$, 
the first sum can be estimated as 
\begin{multline*}
  \sum_{k:\,|k/N-\tfrac{1}{2}| \le \pi/(16 M^2 + 4 \pi^2)^{1/2}} 
2^{-N} {N \choose
    k} \frac{|k/N-\tfrac{1}{2}|}{\sqrt{1-4(k/N-\tfrac{1}{2})^2}} \\ 
\le \frac{2}{\sqrt{3}}
  \sum_{k:\,|k/N-\tfrac{1}{2}| \le \pi/(16 M^2 + 4 \pi^2)^{1/2}} 
2^{-N} {N \choose k}
  \left|\frac{1}{2}-\frac{k}{N}\right|.
\end{multline*}
The second sum can be estimated by 
\begin{multline*}
      \frac{3}{4} \frac{\pi}{M} 
\sum_{k:\,|k/N-\tfrac{1}{2}| > \pi/(16 M^2 +
    4 \pi^2)^{1/2}} 2^{-N} {N \choose 
    k} \frac{(16M^2 + 4 \pi^2)^{1/2}}{\pi}
\left|\frac{k}{N}-\frac{1}{2}\right|\\ 
\le\,3 \sqrt{1+ \frac{\pi^2}{4M^2}} \sum_{k:\,|k/N-\tfrac{1}{2}| > 
\pi/(16 M^2 +
    4 \pi^2)^{1/2}} 2^{-N} {N \choose 
    k} \left|\frac{k}{N}-\frac{1}{2}\right|.
\end{multline*}
Adding the estimates of these two sums we obtain 
$$
e^{\ap}_{\pr_1} (M,\tfrac{8}{\pi^2}) \le 3 \sqrt{1+ \frac{\pi^2}{4M^2}}
   \; \sum_{k=0}^N 2^{-N} {N \choose k} \left| \frac{1}{2} -
    \frac{k}{N} \right|.
$$
The last sum is given by (\ref{eq:zero_err}) and estimated by
(\ref{needed}). Hence 
$$
e^{\ap}_{\pr_1} (M,\tfrac{8}{\pi^2}) \le \sqrt{\frac{3}{2\pi}}
  \sqrt{1+\frac{\pi^2}{4M^2}} \frac{1}{\sqrt{N-1}}\;
  e^{1/(12(N-1))}
$$
which completes the proof.
\end{proof}

In the next theorem we consider the case when $M$ is not
divisible by 4. 

\vspace{12pt}

\begin{WAn4}
  Assume that $M>4$ is not divisible by $4$, and let $p\in(\tfrac{1}{2},
   \tfrac{8}{\pi^2}]$. 
  Then the
  average-probabilistic error of the \textbf{QS} algorithm with
  respect to the measure $\pr_1$ satisfies
\begin{equation*}
  e_{\pr_1}^{\ap}(M,p) \geq
      \frac{\pi}{4M}\left(1-\frac{1}{M}-\frac{1}{\beta}\right)\left( 1
      - 2 \exp \left(-\,\frac{N\pi^2}{(8\beta M)^2}\right) \right)\qquad
      \forall \beta > 1.
\end{equation*}
\end{WAn4}

\begin{proof}
Let $M=4M'+ \tau$ for $\tau \in \{1,2,3\}$. Let, as before,
$\sigma_a=(M/\pi) \arcsin \sqrt{a}$. As in the proof of Theorem
\ref{thm:WA4}, for $|x| <\tfrac{1}{2}$ we have 
\begin{equation*}
      \sigma_{\tfrac{1}{2}+x} = \frac{M}{4} + \frac{M}{\pi} \frac{x}{2
    \sqrt{(1-\xi_x) \xi_x}}\qquad \text{for} \quad \xi_x 
    \in (\tfrac{1}{2}, \tfrac{1}{2}+x)
\end{equation*}
 and $2 \sqrt{(1-\xi_x)\xi_x} \geq \sqrt{1-4x^2}$. Assume additionally
that 
\begin{equation*}
  \frac{M}{\pi} \frac{|x|}{\sqrt{1-4x^2}}\, \le \,\frac{1}{4\beta},
\end{equation*}
which is equivalent to assuming that 
\begin{equation}
\label{eq:xestim}
  |x| \,\le \,\frac{\pi}{(16M^2\beta^2+4\pi^2)^{1/2}}.
\end{equation}
Thus for $x$ satisfying (\ref{eq:xestim}) we have 
\begin{equation*}
  \sigma_{\tfrac{1}{2}+x}=  M' + \frac{\tau}{4}  + \frac{M}{\pi}
  \frac{x\,\theta(x)}{\sqrt{1-4x^2}}\ \ \mbox{with}\ \ \theta(x)\in[0,1],
\end{equation*}
and $\fl{\sigma_{\tfrac{1}{2}+ x}} = M'$ and  
$\cl{\sigma_{\tfrac{1}{2}+ x}} = M'+1$. 

{}From the proof of Corollary \ref{MaxCor} we have
$\mu(\{\cl{\sigma_{a_f}}, \fl{\sigma_{a_f}}\},f) \ge \tfrac{8}{\pi^2}$.
Since $\mu (A, f) \ge p>\tfrac{1}{2}$ 
then either $\cl{\sigma_{a_f}} \in
A$ or $\fl{\sigma_{a_f}} \in A$. We then estimate
\begin{align}
\label{eq:eap_estim}
  e^{\ap}_{\pr_1} (M,\tfrac{8}{\pi^2}) \ge 
\sum_{f \in \B_N} \pr_1(f)\min \big\{
  \big|a_f - \bar a_f\big(\fl{\sigma_{a_f}}\big)\big|, \big|a_f -  \bar
  a_f\big(\cl{\sigma_{a_f}}\big)\big|\big\}. 
\end{align} 

We now estimate the error of the \textbf{QS} algorithm for~$f \in
\B_N$ such that $a_f = \tfrac{1}{2} +x$ for~$x$ satisfying~(\ref{eq:xestim})
and the outcome $j=\cl{\sigma_{a_f}}=M'$ or
$j=\fl{\sigma_{a_f}}=M'+1$.  Denote the outcome by $M' + \kappa$ for
$\kappa \in \{0,1\}$. By Taylor's theorem we have
\begin{multline*}
  \sin^2\left( \frac{M' + \kappa}{M}\, \pi\right) = \sin^2 \left(
  \frac{\pi}{4} + \frac{\pi}{M} ( \kappa - \tfrac{1}{4}\tau) \right)  \\
= \frac{1}{2} + \sin(2 \xi_{\kappa,\tau}) \frac{\pi}{M} (\kappa - 
   \tfrac{1}{4}\tau)\qquad
  \text{for $\xi_{\kappa,\tau} \in [\tfrac{1}{4}\pi, 
   \tfrac{1}{4}\pi + (\pi/M)(\kappa-\tfrac{1}{4}\tau)]$}.
\end{multline*}
Since $\sin(t)\ge 2t/\pi$ for $t\in [0,\pi/2]$,
we have $|\sin(2 \xi_{\kappa,\tau})| \ge 1- |4 \kappa -\tau|/M$.  
Consider the error for the outcome $M' + \kappa$ and $x$ 
satisfying~(\ref{eq:xestim}). Then $|x| \le \pi/(4\beta M)$
and the error can be estimated by
\begin{align*}
  \left| \frac{1}{2}+ x - \sin^2\left( \frac{M' + \kappa}{M}\,
  \pi\right)\right| &= \left| x - \sin(2 \xi _{\kappa,\tau})
  \frac{\pi}{M} (\kappa - \tfrac{1}{4}\tau)\right| \\
  & \geq \frac{\pi}{M} | \kappa - \tfrac{1}{4}\tau| 
    |\sin(2 \xi_{\kappa,\tau})| -  |x|\\ 
&\geq  \frac{\pi}{4} \frac{|4\kappa-\tau|}{M} \left( 1 - \frac{|4\kappa -\tau |}{M}
  \right) - \frac{\pi}{4\beta M}.
\end{align*}
Clearly, $|4\kappa-\tau|\in\{1,2,3\}$ and $|4\kappa-\tau|/M\in[1/M,3/M]$.
Then $|4\kappa-\tau|(1-|4\kappa-\tau|/M)/M\ge (1-1/M)/M$. Therefore
\begin{align*}
  \left| \frac{1}{2}+ x - \sin^2\left( \frac{M' + \kappa}{M}\,
  \pi\right)\right| \geq 
\frac{\pi}{M} \left( \frac{1}{4} \left( 1 - \frac{1}{M} \right)
  - \frac{1}{4 \beta} \right) = \frac{\pi}{4M}\left(1-\frac{1}{M} - \frac{1}{\beta}\right).
\end{align*}
Hence, for $f$ such that $a_f = \tfrac{1}{2} +x $ with $x$ satisfying (\ref{eq:xestim})
we have 
\begin{align}
\label{eq:123}
   \min \big\{ \big|a_f - \bar a_f\big(\fl{\sigma_{a_f}}\big)\big|,
  \big|a_f - \bar a_f\big(\cl{\sigma_{a_f}}\big)\big|\big\} \ge
  \frac{\pi}{4M}\left(1-\frac{1}{M} - \frac{1}{\beta}\right).
\end{align}

We are now ready to estimate $e_{\pr_1}^{\ap} (M, p)$. First,
by (\ref{eq:eap_estim}), we have 
\begin{align*}
  e^{\ap}_{\pr_1} (M, p) \ge \sum_{f:\,|a_f - \tfrac{1}{2}| 
\le (16M^2\beta^2
  + 4 \pi^2)^{1/2}} \pr_1(f) \min \big\{
  \big|a_f - \bar a_f\big(\fl{\sigma_{a_f}}\big)\big|, \big|a_f -  \bar
  a_f\big(\cl{\sigma_{a_f}}\big)\big|\big\}.
\end{align*}
This,  (\ref{eq:123}) and the Bernstein inequality, 
$
\sum_{k:\,|k/N-\tfrac{1}{2}| > \ve} 2^{-N} {N \choose k} \le 2 e^{-N \ve^2 /4},
$ 
yields
\begin{align*}
  e^{\ap}_{\pr_1} (M, p) & \ge \sum_{f:\,|a_f - \tfrac{1}{2}| \le
    \pi(16M^2\beta^2
    + 4 \pi^2)^{1/2}}\pr_1(f) \,\frac{\pi}{4M}\left(1-\frac{1}{M} - \frac{1}{\beta}\right) \\
  &=\frac{\pi}{4M}\left(1-\frac{1}{M} - \frac{1}{\beta}\right)
    \sum_{k:\,|k/N - \tfrac{1}{2}|\le \pi(16M^2\beta^2 
    + 4 \pi^2)^{1/2}} 2^{-N} {N \choose k} \\
  &\geq \frac{\pi}{4M}\left(1-\frac{1}{M} - \frac{1}{\beta}\right) \left( 1 - 2 \exp \left(- \frac{N\pi^2}{4( 16
        M^2 \beta^2  - 4)}\right) \right) \\
  &\geq \frac{\pi}{4M}\left(1-\frac{1}{M} - \frac{1}{\beta}\right) \left( 1 - 2 \exp
    \left(-\frac{N\pi^2}{(8 \beta M)^2}\right) \right), 
\end{align*}
which completes the proof.
\end{proof}

\vspace{12pt}

Obviously, in the average-probabilistic setting, we should use 
the \textbf{QS} algorithm with $M$ divisible by $4$. Then Theorem
\ref{thm:WA4} states that the error is of order $\min\{M^{-1},N^{-1/2}\}$.
Recently, Papageorgiou \cite{Papa} proved that for any quantum 
algorithm that uses $M$ quantum queries the error 
is bounded from below by of $c\min\{M^{-1},N^{-1/2}\}$
with probability $p\in(\tfrac{1}{2},\tfrac{8}{\pi^2}]$. 
Here, $c$ is a positive number independent of $M$ and $N$. 
Hence, the \textbf{QS} algorithm
enjoys an optimality property also in the average-probabilistic setting
for the measure $\pr_1$ as long as we use it with $M$ divisible by $4$.

We now turn to the measure $\pr_2$. Clearly, the average-probabilistic 
error of the \textbf{QS} algorithm is bounded by its worst-probabilistic
error, which is of order $M^{-1}$ with probability $p\in(\tfrac{1}{2},
\tfrac{8}{\pi^2}]$. 
It turns out, again due to a recent result of Papageorgiou \cite{Papa}
that this bound is the best possible, since any quantum algorithm that uses
$M$ quantum queries must have an error proportional at least to $M^{-1}$.
Hence, the factor $N^{-1/2}$ that is present
for the measure $\pr_1$ does not appear for the measure 
$\pr_2$, and the behavior of the \textbf{QS} algorithm
is roughly the same in the worst- and average-probabilistic settings. 

\section*{Acknowledgment}

We are grateful for many valuable comments on this paper
from J. Creutzig, A. Papageorgiou, J. F. Traub, and A.G. Werschulz.


\begin{thebibliography}{99}
  
\bibitem{BMHT} G. Brassard, P. H\o{}yer, M. Mosca, A. Tapp, Quantum
  Amplitude Estimation and Amplification,
  http://arXiv.org/quant-ph/0005055, 2000. 

\bibitem{G98} L. Grover, A fast quantum mechanical algorithm for
database search, Proceedings of the 28th Annual ACM Symposium on the Theory
of Computing  (STOC), 212-219, 1996, http://arXiv.org/quant-ph/9605043, 1996.


\bibitem{NC} M. A. Nielsen, I. L. Chuang \emph{Quantum Computation and
    Quantum Information }, Cambridge University Press, 2001,


\bibitem{H1} S. Heinrich, Quantum Summation with an Application to
       Integration, \emph{J. Complexity}, \textbf{18}, 1-50, 2002,  
http://arXiv.org/quant-ph/0105116, 2001. 

\bibitem{H2} S. Heinrich, From Monte Carlo to Quantum Computation,
http://arXiv.org/quant-ph/0112152, 2001.

\bibitem{HN} S. Heinrich and E. Novak, On a Problem in Quantum
  Summation, to appear in \emph{J. Complexity}, \textbf{19}, 2003,
  http://arXiv.org/quant-ph/0109038, 2001.

  
\bibitem{NayakWu} A. Nayak and F. Wu, The quantum query complexity of
  approximating the median and related statistics, Proceedings of the 31th
  Annual ACM Symposium on the Theory of Computing (STOC), 384-393, 1999,
  http://arXiv.org/quant-ph/9804066, 1998.

\bibitem{N00} E. Novak, Quantum Complexity of Integration,
 \emph{J. Complexity}, \textbf{17}, 2-16, 2001,
http://arXiv.org/quant-ph/0008124, 2000.

\bibitem{NSW} E. Novak, I. H. Sloan and H. Wo\'zniakowski,
Tractability of Approximation for Weighted Korobov Spaces on classical
and Quantum Computers, http://arXiv.org/quant-ph/0206023, 2002.


\bibitem{Papa} A. Papageorgiou, Average case quantum lower bounds
for computing the Boolean means, in progress. 

\bibitem{TW} J. F. Traub and H. Wo\'zniakowski, Path Integration on a
Quantum Computer, to appear in
\emph{Quantum Information Processing},
http://arXiv.org/quant-ph/0109113, 2001.

\end{thebibliography}
\end{document}